\documentclass[preprint, 3p]{elsarticle}
\journal{CAS}
\usepackage[doublespacing]{setspace}
\usepackage{xspace}
\usepackage[utf8]{inputenc}
\usepackage{subcaption}
\usepackage[export]{adjustbox}
\usepackage{amsmath,bm}
\usepackage{amssymb}
\usepackage{tikz}
\usetikzlibrary{arrows, shapes.geometric}
\usepackage{siunitx}
\usepackage{graphicx}
\usepackage{tabularx}
\usepackage{pstool}
\usepackage{booktabs}
\usepackage{color}
\usepackage{fancyhdr}
\usepackage{pgfplots}
\pgfplotsset{compat=1.12}
\usepackage{circuitikz}
\newcommand{\nc}{\newcommand}
\nc{\rnc}{\renewcommand}
\nc{\bs}{\boldsymbol}
\nc{\RM}[1]{\MakeUppercase{\romannumeral #1}}
\nc{\define}{:=}
\nc{\ul}{\underline}
\nc{\mm}{\boldsymbol}
\nc{\mms}{\mm}
\nc{\real}[1]{\operatorname{Re}\left\lbrace #1 \right\rbrace}
\nc{\imag}[1]{\operatorname{Im}\lbrace #1 \rbrace}
\nc{\abs}[1]{\left| #1 \right|}
\nc{\redcol}[1]{\textcolor{red}{#1}}
\nc{\ie}{i.e.\,}
\nc{\eg}{e.g.\,}
\nc{\etc}{etc.\,}
\nc{\cf}{c.\,f.\,}
\nc{\etal}{et\,al.\,}
\nc{\sref}[1]{Section \ref{sec:#1}}
\nc{\aref}[1]{\ref{append:#1}}
\nc{\eref}[1]{Eq.\ (\ref{eq:#1})}
\nc{\fref}[1]{Fig.~\ref{fig:#1}}
\nc{\tref}[1]{Tab.\ \ref{tab:#1}}
\nc{\fk}{\,,}
\nc{\fp}{\,.}
\nc{\herm}{{}^{\mathrm H}}
\nc{\tra}{{}^{\mathrm T}}
\rnc{\matrix}[2]{\left[\!\!\begin{array}{#1} #2\end{array}\!\!\right]}
\rnc{\vector}[1]{\matrix{c}{#1}}
\nc{\ee}{\mathrm{e}}
\nc{\ii}{\mathrm{i}}
\nc{\dd}{\mathrm{d}}
\nc{\e}[2]{\begin{equation} #1 \label {eq:#2} \end{equation}}
\nc{\ea}[2]{
\begin{eqnarray}
#1 \label {eq:#2} \end{eqnarray}}
\nc{\east}[1]{
\begin{eqnarray}
#1 
\end{eqnarray}}
\nc{\fig}[4][tbh]{\begin{figure}[#1]
\centering
\includegraphics[width=#4\linewidth]{figs/#2}\caption{#3\label{fig:#2}}
\end{figure}}

\nc{\usensor}{\hat{u}_{\mathrm{Sensor}}}
\nc{\umodal}{\hat{u}}
\nc{\usolid}{\hat{\mm u}^\mathrm{s}}
\nc{\usolidtd}{\mm u^\mathrm{s}}
\nc{\usolidm}{\mm q}
\nc{\usolidtddot}{\dot{\mm  u}^\mathrm{s}}
\nc{\ufluid}{\hat{\mm u}^\mathrm{a}}
\nc{\ufluidtd}{\mm u^\mathrm{a}}
\nc{\forcec}{\hat{\mm f}^\mathrm{c}}
\nc{\forcea}{\hat{\mm f}^\mathrm{a}}
\nc{\forcectd}{\mm f^\mathrm{c}}
\nc{\forceatd}{\mm f^\mathrm{a}}
\nc{\Dsnma}{D^{\mathrm{s}}(\omega^{\mathrm{nma}},\mm \Psi^\mathrm{nma})}
\nc{\Danma}{D^{\mathrm{a}}(\omega^{\mathrm{nma}},\mm \Psi^\mathrm{nma})}
\nc{\Dsnmaair}{D^{\mathrm{s}}(\omega^{\mathrm{nma,Air}},\mm \Psi^\mathrm{nma,Air})}
\nc{\Danmaair}{D^{\mathrm{a}}(\omega^{\mathrm{nma,Air}},\mm \Psi^\mathrm{nma,Air})}
\nc{\Dslin}{D^{\mathrm{s}}(\omega^{\mathrm{stick}},\mm \Psi^{\mathrm{stick}})}
\nc{\Dalin}{D^{\mathrm{a}}(\omega^{\mathrm{stick}},\mm \Psi^{\mathrm{stick}})}
\nc{\Danmaconstpsi}{D^{\mathrm{a}}(\omega^{\mathrm{nma}},\mm \Psi^{\mathrm{stick}})}
\nc{\Danmaconstom}{D^{\mathrm{a}}(\omega^{\mathrm{stick}},\mm \Psi^{\mathrm{nma}})}
\nc{\fqref}{f_\mathrm{ref}}

\nc{\ua}{\mm u^{\mathrm a}}
\nc{\uhs}{\hat{\mm u}_{\mathrm s}}
\nc{\uha}{\hat{\mm u}_{\mathrm a}}
\nc{\uhsh}{\hat{\mm u}^{\mathrm s}_h}
\nc{\uhsone}{\hat{\mm u}^{\mathrm s}_1}
\nc{\uhshs}{\hat{\mm u}^{\mathrm s}_{\Hs}}
\nc{\Hs}{H^{\mathrm s}}
\nc{\Ha}{H^{\mathrm a}}
\nc{\EO}{EO}
\nc{\rpc}{r_{\mathrm{pc}}}

\makeindex             % used for the subject index
\begin{document}
\begin{frontmatter}
\title{Fully Coupled Forced Response Analysis of Nonlinear Turbine Blade Vibrations in the Frequency Domain}

\renewcommand{\thefootnote}{\fnsymbol{footnote}}
%\author[1]{Christian Berthold\footnote{Corresponding author. Email: christian.berthold@dlr.de}}
\author[1]{Christian Berthold\footnote{Corresponding author.}}
\author[2]{Johann Gross}
\address[1]{German Aerospace Center, Linder Hoehe 1, 51147 Cologne, Germany}
\author[1]{Christian Frey}
\author[2]{Malte Krack}
\address[2]{University of Stuttgart, Pfaffenwaldring 6, 70569 Stuttgart, Germany}

\begin{abstract}
For the first time, a fully-coupled Harmonic Balance method is developed for the forced response of turbomachinery blades.
The method is applied to a state-of-the-art model of a turbine bladed disk with interlocked shrouds subjected to wake-induced loading.
The recurrent opening and closing of the pre-loaded shroud contact causes a softening effect, leading to turning points in the amplitude-frequency curve near resonance.
Therefore, the coupled solver is embedded into a numerical path continuation framework.
Two variants are developed:
the coupled continuation of the solution path, and
the coupled re-iteration of selected solution points.
While the re-iteration variant is slightly more costly per solution point, it has the important advantage that it can be run completely in parallel, which substantially reduces the wall clock time.
It is shown that wake- and vibration-induced flow fields do not linearly superimpose, leading to a severe underestimation of the resonant vibration level by the influence-coefficient-based state-of-the-art methods (which rely on this linearity assumption).
\end{abstract}

\end{frontmatter}

\renewcommand{\thefootnote}{\arabic{footnote}}
\section{Introduction}
% TECHNICAL RELEVANCE OF BLADE VIBRATIONS; MOST IMPORTANT VIBRATION MECHANISMS
The most important vibration mechanisms of turbomachinery blades are resonant excitation due to rotation through the circumferentially inhomogeneous pressure field (\emph{Forced Response}), and self-excitation due to unstable mutual aerodynamic interaction among the blades within a cascade (\emph{Flutter}) \cite{srin1997}.
% DESIGN AGAINST VIBRATIONS
The current strategy to deal with these vibration mechanisms in engineering design is as follows:
First, flutter must be avoided. % by requiring positive total damping.
Secondly, resonances with the lowest natural frequencies (approx. 5-6 mode families) in the operating range must be avoided.
Finally, the remaining resonances must be endured without high-cycle fatigue.
Flutter avoidance has become a dominant constraint, especially in low-pressure turbines, \ie the blades cannot be further improved without causing flutter \cite{Waite.2014}.
Therefore, the concept of flutter-tolerant design is currently explored \cite{petr2012c,Lassalle.2018b,Ombret.2021}.
\\
% DAMPING; incl. INHERENTLY NONLINEAR FRICTION DAMPING
To ensure a vibration-safe design, the prediction of the total damping is crucial, which is composed of aerodynamic and structural mechanical damping.
To avoid flutter, a positive total damping ratio is required.
The total damping ratio also determines how large the vibration stresses become in the resonant case.
% The aerodynamic damping depends on the (reduced) frequency and the vibration mode.
The damping of the bulk material is almost always negligible, so that the structural mechanical damping is dominated by the dissipative dry frictional interactions occurring, for instance, at interlocked shrouds, under-platform dampers, or the blade-disk attachment.
The friction contact interactions pose a significant challenge in the prediction, as they can only be described using nonlinear force-displacement relationships.
The contact interactions also determine the effective stiffness of the joints and lead to an amplitude dependence of the natural frequencies.
\\
% STATE OF THE ART VIBRATION PREDICTION
For the calculation of periodic oscillations in turbomachinery, \emph{Harmonic Balance (HB)} is commonly used, both in aerodynamics/aero-elasticity \cite{Hall.2002} and in structural dynamics \cite{Krack.2016}.
Here, the generalized coordinates or the conservative variables are sought in the form of a truncated Fourier series and the differential equations governing the solid and fluid dynamics are transformed into the frequency domain.
The method is able to account for nonlinearities such as the contact interactions in joints as well as strongly unsteady phenomena due to flow separation and shocks.
While nonlinear dynamic aero-elasticity could in principle be analyzed using conventional multi-physics tools based on time step integration \cite{Li.2018d}, Harmonic Balance reduces the computation effort by several orders of magnitude, making computations on high-quality models feasible.
Reasons for the high numerical efficiency are the bypassing of the usually very long transient, the compact representation of the oscillation with few (resonant) Fourier terms (instead of time levels), avoidance of numerical instability/damping (typical for time step integration applied to stiff differential equations due to turbulence and contact models), high quality implementation of non-reflecting boundary conditions, as well as the natural formulation of time-delay boundary conditions.
Here, the time-delay boundary conditions allow the restriction to only one reference sector (a blade-disk segment or passage) of the rotational periodic system, which is a considerable simplification for the typical blade counts in axial turbomachinery.
\\
% FULLY-COUPLED HB METHOD vs. PREVAILING PRACTICE (AERODYNAMIC FORCES = IMPOSED FORCES + LINEARIZED VIBRATION-INDUCED FORCES; AIC)
It is the prevailing practice to approximate the aerodynamic forces as a superposition of imposed forces (which would occur for rigid, rotating blades) and (elastic) vibration-induced forces \cite{Carstens.2003}.
The latter are linearized and expressed by means of modal aerodynamic influence coefficients (AIC).
Usually only one AIC is determined with respect to a single mode, see \eg \cite{Lassalle.2018b,Ombret.2021}, in a few exceptions a matrix is determined with respect to a small number of modes.
In any case, the contact boundary conditions are linearized for the computation of the modes (usually assuming sticking contact).
The described procedure neglects the amplitude dependence of the modal deflection shape (due to nonlinear contact interactions).
As shown in \cite{Berthold.2020,Berthold.2021} for state-of-the-art models of a low-pressure turbine blade row with interlocked tip shrouds, this can lead to a considerable over- or underestimation of the aerodynamic damping and the resulting vibration behavior.
In fact, the aerodynamic damping may change its sign from positive to negative, or vice versa, under consideration of the amplitude-variable modal deflection shape.
As a consequence, the stability limit, beyond which friction damping cannot stabilize flutter-induced vibrations, may vanish.
Or perhaps more critically, the phenomenon of nonlinear instability may occur, where the equilibrium state (no vibrations) is stable according to linear theory, but a small perturbation (inevitable in reality) may lead to divergence (unbounded growth of vibrations).
The above described prevailing practice also neglects the amplitude-dependence of the aerodynamic forces.
This effect was relatively small in the aforementioned studies \cite{Berthold.2020,Berthold.2021}.
It will be shown to have a considerable effect in the present study.
Finally, it is a common practice to neglect the frequency dependence of the aerodynamic influence, which may have an effect of similar magnitude as the mode shape variation \cite{Berthold.2020,Berthold.2021}.
\\
% PURPOSE OF PRESENT WORK (see also GTP_Bidiko_ForcedResponse.tex)
The fully coupled method developed in \cite{Berthold.2021} is based on Harmonic Balance and overcomes all previously mentioned deficiencies.
However, it is so far limited to flutter-induced friction-damped limit cycle oscillations.
In the present work, the method is further developed and applied to the analysis of forced response.
This is accompanied with two important challenges:
First, it is crucial to analyze how the vibrations vary with a parameter (the rotor speed) in order to capture the resonant behavior.
To this end, the method will be combined with numerical path continuation.
Second, the unsteady flow field is not only induced by vibration but has already a substantial unsteady part in the absence of vibration, which is responsible for the dynamic excitation forces.
In the numerical example, the latter is caused by wakes from the upstream cascade.
It will be shown that the wake-induced and the vibration-induced aerodynamic forces do not linearly superimpose, which corresponds to another form of nonlinear fluid-structure interaction (not relevant in the aforementioned studies).
Next, the aero-elastic problem setting is described, and different variants of the coupled forced response analysis are proposed.
These are assessed for the considered numerical example, and conclusions are drawn.
\section{Aero-elastic Problem Setting}
% GENERAL PROBLEM SETTING AND SYSTEM BOUNDARIES
A state-of-the-art model of a low pressure turbine rotor stage subjected to aerodynamic wake excitation is considered as numerical example in the present work.
The bladed disk illustrated in \fref{T16001} comprises 60 blades with interlocked tip shrouds.
The steady-state forced response is sought in the form of a symmetric and time-periodic traveling wave.
This permits to reduce the problem domain to a single sector, containing one section of the bladed disk and one passage, with appropriate time-delay boundary conditions.
This is a common idealization in the presence of strong inter-sector coupling, here via tip shrouds, because then mistuning-induced localization and amplitude magnification are usually negligible \cite{wei1988b}.
A partitioned modeling approach is pursued, where the solid is described using three-dimensional finite elements, and the flow is resolved using a finite volume method.
The time-periodic solution of the governing differential equations in both domains is obtained using Harmonic Balance.
In the following subsections, the models of the structure and the flow are further described.
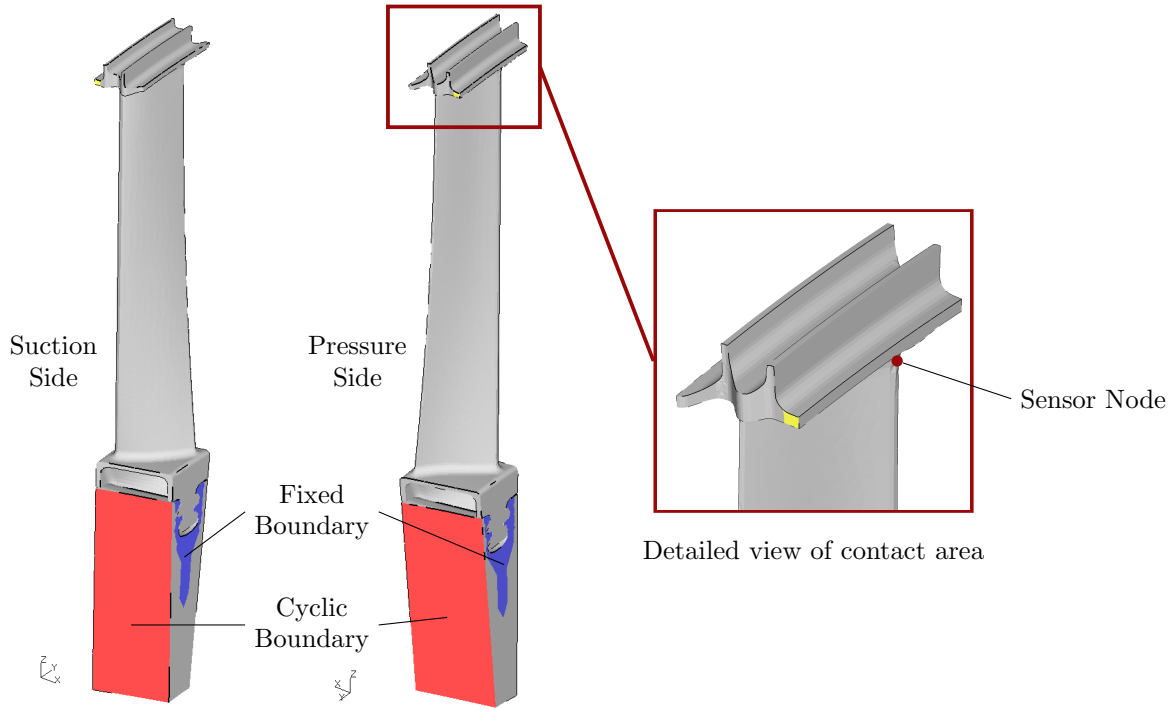
\begin{figure}[t]
    \centering
        \begin{tikzpicture}[>=latex]
        \definecolor{darkred}{HTML}{990808}
        \centering
            \node[inner sep=0pt] (bladess) at (-2,0) {\includegraphics[width=.155\textwidth]{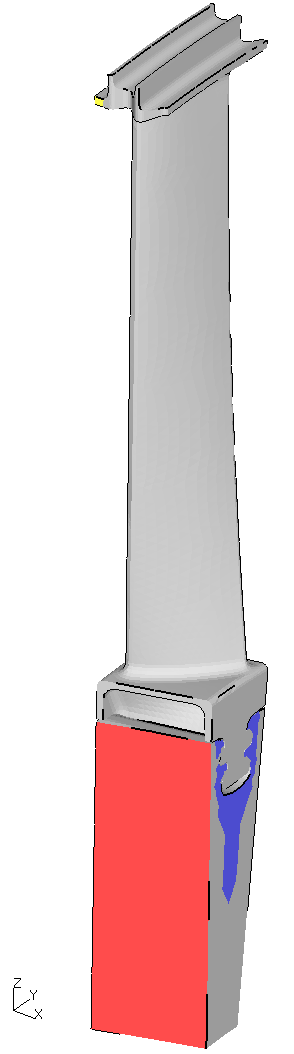}};
            \node[left of=bladess, align=center] () {Suction \\ Side};
            \node[inner sep=0pt] (bladeps) at (2,0) {\includegraphics[width=.17\textwidth]{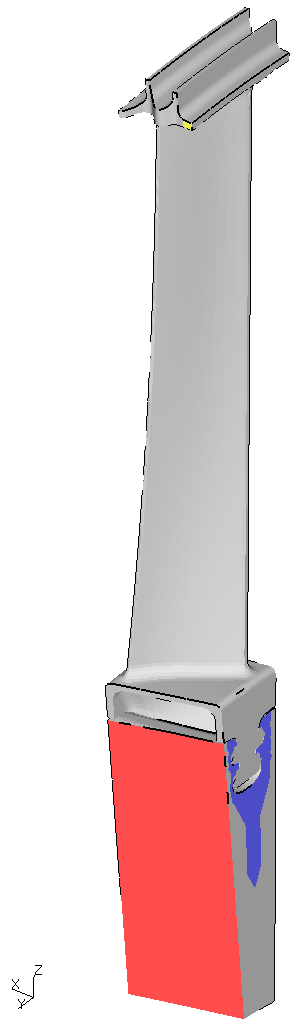}};
            \node[left of=bladeps, align=center] () {Pressure \\ Side};
            \node[inner sep=0pt, line width=1.0mm, draw=darkred] (bladeshroud) at (7,0) {\includegraphics[width=.25\textwidth]{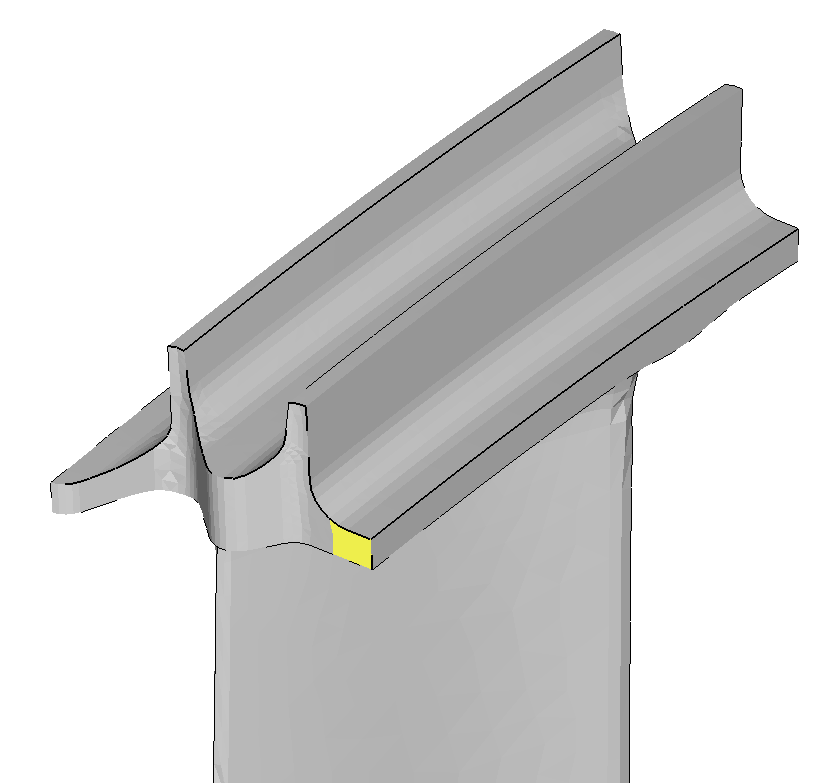}};
            \node[below of=bladeshroud, align=center, yshift=-1.5cm] () {Detailed view of contact area};
            \node[circle,fill=darkred,inner sep=1.5pt](sensornode) at (8.1,0.0){};
            \draw (sensornode) -- +(1.5,-0.5) node[right](){Sensor Node};
            \draw[draw=darkred, line width=0.5mm] (1.4,4.7) rectangle ++(2,-1.6);
            \draw[draw=darkred, line width=0.5mm] (3.4, 4.7-1.6/2) -- (bladeshroud.west);

            \node[align=center](fixed boundary) at (0.4,-2.0){Fixed\\Boundary};
            \draw (fixed boundary.east) -- +(1.65,-0.7);
            \draw (fixed boundary.west) -- +(-0.8,-0.6);
            
            \node[align=center](cyclic boundary) at (0.4,-3.5){Cyclic\\Boundary};
            \draw (cyclic boundary.east) -- +(0.9,0.1);
            \draw (cyclic boundary.west) -- +(-1.5,0.0);
        \end{tikzpicture}
    \caption{Structural model of the reference sector of the bladed disk of the considered low pressure turbine. The detailed view illustrates the contact area (yellow).}
    \label{fig:T16001}
\end{figure}
\subsection{Structure model}
The model is largely adopted from \cite{Berthold.2020}, where additional details are given \eg on the meshing and the nonlinear static analysis carried out to account for the centrifugal loading and the subsequent linearization.
Nonlinear contact interactions are considered in the yellow area indicated in \fref{T16001}.
Unilateral elastic interactions are accounted for in the normal contact direction and elastic dry friction is modeled in the tangential contact plane.
The friction coefficient is set to $0.5$ and the contact stiffness per area is specified as $5.3563\times10^4~\mathrm{N}/\mathrm{mm}^3$.
The contact laws are imposed using contact elements coinciding with the faces of the underlying solid elements.
The 12 node pairs of the conforming contact mesh are used as integration points.
The shroud contact and the aerodynamic loading are the only considered sources of nonlinearity.
This permits to reduce the linear finite element model of the inner structure using the conventional Craig-Bampton method.
More specifically, the static constraint modes associated with the three relative displacements at each contact node pair are retained, along with the 30 lowest-frequency fixed-interface normal modes (based on a preliminary convergence study).
Thus, the reduced structural sector model contains a total of $M=66$ component modes.
According to Harmonic Balance, the corresponding generalized coordinates, $\mm u^{\mathrm s}\in\mathbb R^{M\times 1}$, are sought in the form of a Fourier series,
\east{
\mm u^{\mathrm s} = \real{\sum\limits_{h=0}^{\Hs}\uhsh\ee^{\ii\, h\, \EO\, \Omega \, t}}\fk \label{eq:Fourier}
}
truncated to order $\Hs$.
Herein, $\uhsh$ are the complex Fourier coefficients, $\ii=\sqrt{-1}$ is the imaginary unit, $\EO$ is the engine order of the aerodynamic wake excitation, $\Omega$ is the rotational velocity, and $t$ denotes time.
\\
The Harmonic Balance equations of the reduced structural sector model read
\east{
\mm S_h(\Omega)\uhsh + \hat{\mm g}_h\left(\uhs\right) - \hat{\mm f}_h &=& \mm 0 \quad h=0,1,\ldots,\Hs\fp\label{eq:HBs} \\
}
Herein, $\mm S_h$ is the dynamic stiffness matrix associated with linear elastic, material damping and inertia forces, and $\hat{\mm g}_h$ and $\hat{\mm f}_h$ are the Fourier coefficients of the contact and the aerodynamic forces, respectively.
Due to the nonlinearity, $\hat{\mm g}_h$ depend on all Fourier coefficients of the generalized coordinates, stacked in the vector $\uhs = [\uhsone;\ldots;\uhshs]$.
These are evaluated in discrete time using the conventional alternating-frequency-time scheme, see \eg \cite{Krack.2016}.
$\hat{\mm f}_h$ depend on the aerodynamic variables, as described in the next subsection.

\subsection{Flow model}
% GOVERNING EQUATIONS
The compressible unsteady Reynolds-averaged-Navier-Stokes (URANS) equations are formulated with respect to the conservative variables (density, linear momentum in three directions, energy) and with respect to the rotor's rotating frame of reference.
The equations are closed with the $k$-$\omega$ turbulence model \cite{Wilcox1988}, the ideal gas law and the Sutherland law for the molecular viscosity.
The passage is discretized with a finite volume mesh containing $C= 476,000$ cells.
The aerodynamic state variables are represented as Fourier series analogous to \eref{Fourier}, but generally with a different truncation order $\Ha$ (instead of $\Hs$).
For each cell $j$, Harmonic Balance yields the equation \cite{Ashcroft.2014}:
\east{
\ii h \EO \Omega\left(\widehat{V\mm u_j^{\mathrm a}}\right)_h + \hat{\mm\rho}_{jh}\left(\uha,\Omega,\hat{\mm\chi}\right) &=& \mm 0 \quad h=0,1,\ldots,\Ha \quad j=1,\ldots,C\fp \label{eq:HBa}
}
Herein, $\left(\widehat{V\mm u_j^{\mathrm a}}\right)_h$ are the Fourier coefficients of the conservative variables multiplied with the respective cell volume.
This formulation is the result of the Arbitrary Eulerian-Lagrangian (ALE) formulation of the Navier-Stokes equation in integral form which is suitable for moving meshes.
$\hat{\mm\rho}_{jh}$ is the $h$-th Fourier coefficient of the nonlinear terms in the URANS equations, associated with the mass-, momentum-, and energy-flux-balance for the finite volume cell $j$.
$\uha$ and $\hat{\mm\chi}$ are set up as $\uhs$, where $\hat{\mm\chi}$ contains the Fourier coefficients of the coordinates of the finite volume mesh.
\\
% MOVING MESH
At run time, the coordinates of the finite volume mesh are determined as the linear combination
\e{ \mm \chi = \mm \chi_0 +  [ \mm \chi_1\, \mm \chi_2 \cdots \mm\chi_M] \mm u^{\mathrm s}\fk}{deformedmeshsuperposition}
where $\mm \chi_0$ represents the undeformed mesh, and $\mm \chi_m$ with $m\geq 1$ the mesh deformation for unit displacement of generalized coordinate $m$.
The latter are pre-computed for each of the $M$ component modes, by solving a Laplace equation, facilitating a simple evaluation of \eref{deformedmeshsuperposition} at run time.
Note that the generalized coordinates all contribute to the movement of the fluid-structure interface.
At the blade surface, the nodal displacements are mapped on the (non-conforming) fluid mesh via a bilinear interpolation.
The described mesh deformation procedure preserved good mesh quality throughout the amplitude range considered in this work.
\\
% AERODYNAMIC FORCE COMPUTATION
By integration the aerodynamic pressure over the fluid mesh surface, consistent nodal forces are obtained.
These are projected onto the component modes to obtain the generalized aerodynamic forces.
\\
% INLET/OUTLET BOUNDARY CONDITIONS incl. WAKE; STEADY FLOW CONDITIONS
At the inlet and the exit, two-dimensional non-reflecting boundary conditions are imposed \cite{Schluess2018}.
The Mach number in the relative frame of reference at the inlet is \num{0.33} and at the exit \num{0.75}.
The Mach contour at 80\% channel height is depicted in \fref{mach}.
The Reynolds number with respect to the exit flow conditions and the chord of the blade is \num{720000} and the stagnation pressure ratio in the absolute frame of reference amounts to \num{1.17}.
\fig{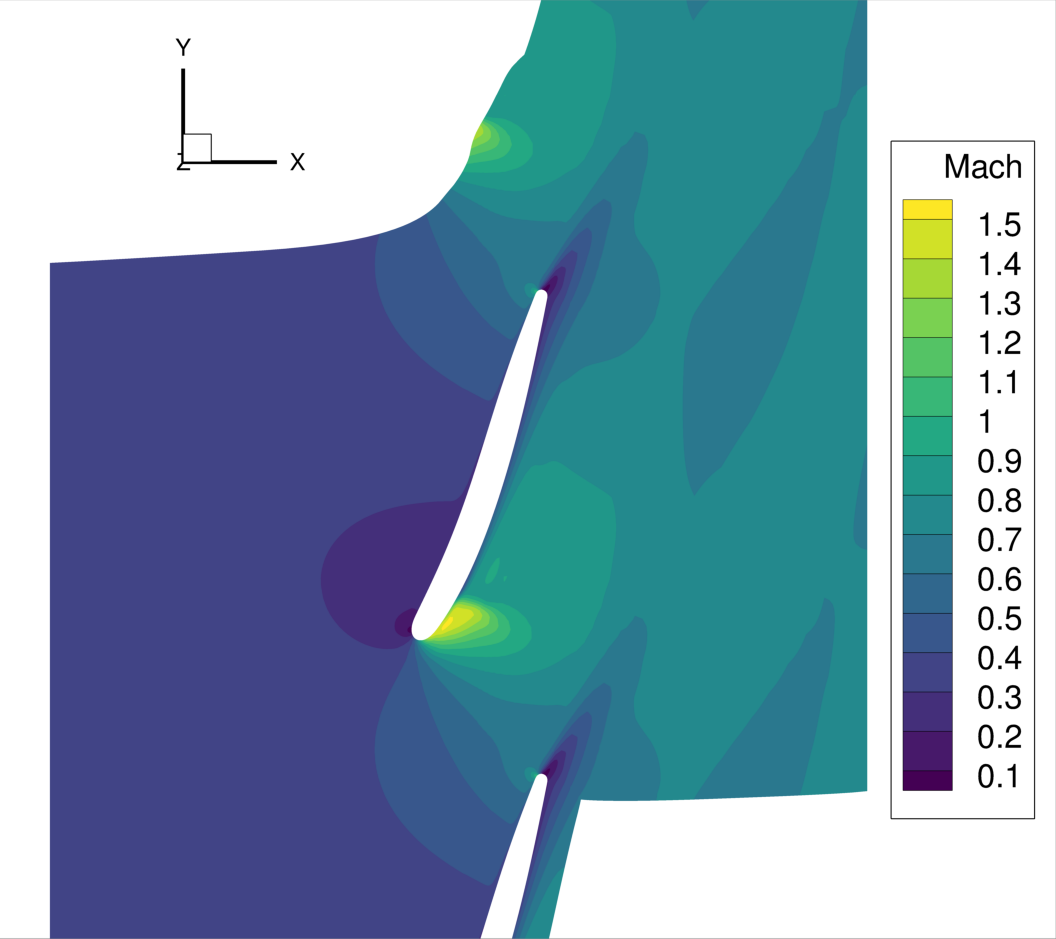}{Mach number at 80\% channel height for the steady flow field.}{0.7}
The aerodynamic wake excitation is modeled as an imposed traveling-wave type inlet disturbance, corresponding to an Engine Order of $\EO=20$.
The disturbance resembles an acoustic and a vorticity wave.
The prescribed generic wake is illustrated in \fref{seg6stagP} for six segments/blades in terms of the stagnation pressure, normalized by the maximum stagnation pressure at the inlet.
The three blades on the right side are plotted transparently while the remaining blades on the left side are removed for better visibility of the inlet plane.

\fig{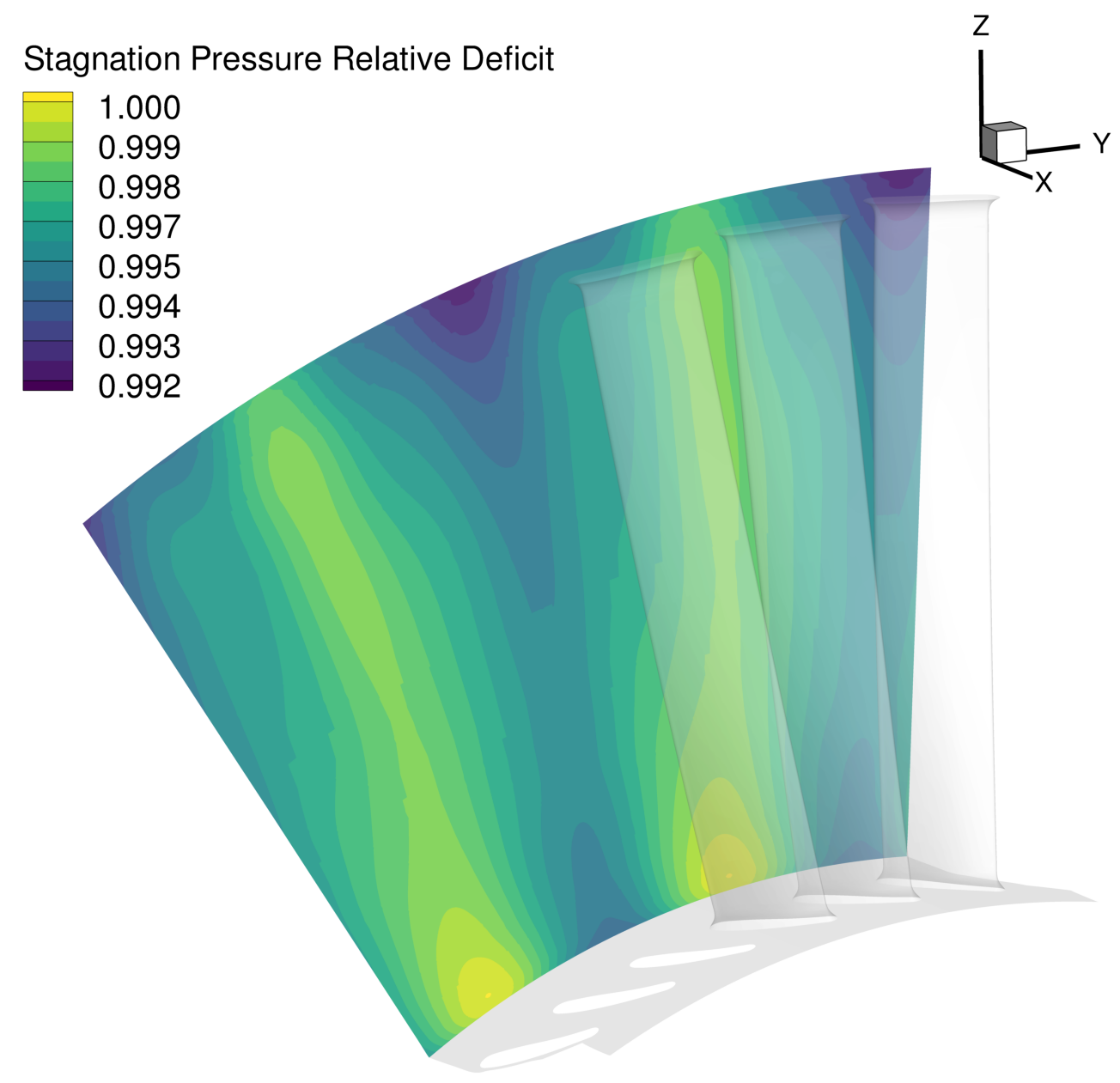}{Inlet stagnation pressure relative deficit contour of generic wake excitation}{0.65}
\section{Coupled Forced Response Analysis}
% JUSTIFICATION OF PARTITIONED APPROACH
The \emph{partitioned coupling approach} pursued in the present work has several advantages:
It ensures consistent modeling with decoupled approaches, a maximum reuse of existing code, and the use of problem-adapted numerical methods in the respective domains.
Perhaps most importantly, it permits using different solvers.
Due to the high number of unknowns in the fluid domain, practically only a pseudo-time solver can be applied.
This has the important downside that it can only compute asymptotically stable limit states.
Thus, a monolithic approach, based on a pseudo-time solver, is not capable of computing, for instance, stability limits in the case of flutter \cite{Berthold.2021} or the unstable solutions on the overhanging forced response branch shown later.
In the partitioned approach proposed in this work, a conventional Newton-type solver is applied in the structural domain, overcoming the above described limitation.
\\
% NEED FOR NUMERICAL PATH CONTINUATION
The occurrence of overhanging branches / turning points necessitates the use of \emph{numerical path continuation}.
To this end, the rotor speed / excitation frequency parameter, $\Omega$, is treated as an additional unknown.
Also, the equation system is augmented by a constraint equation, $\rpc=0$, to ensure local uniqueness and define where on the solution branch the next point ends up (\emph{parametrization constraint}).
In addition, path continuation facilitates numerical efficiency and robustness in ranges with strong gradient changes of the solution path, which are typical near resonances.
It is useful to recall that the exact location of the resonance peak is a priori unknown and may significantly deviate from the natural frequency of the linearized system in the nonlinear case.
\\
% RECAP OF PROBLEM SETTING
The governing equations of the coupled problem can be summarized as
\east{
\mm r_{\mathrm s}\left(\mm x_{\mathrm s},\Omega,\hat{\mm f}\left(\mm x_{\mathrm a}\right)\right) &=& \mm 0\fk \label{eq:rs} \\
\mm r_{\mathrm a}\left(\mm x_{\mathrm a},\Omega,\hat{\mm\chi}\left(\mm x_{\mathrm s}\right)\right) &=& \mm 0\fk \label{eq:ra} \\
r_{\mathrm{pc}}\left(\mm x_{\mathrm s},\mm x_{\mathrm a},\Omega\right) &=& 0\fp \label{eq:rpc}
}
The residual vector functions $\mm r_{\mathrm s}$ and $\mm r_{\mathrm a}$ correspond to the Harmonic Balance equations for the structural and fluid domain, \eref{HBs} and \eref{HBa}, respectively.
The problem has been cast into real arithmetic; \ie, $\mm r_{\mathrm s}$ and $\mm r_{\mathrm a}$ are real-valued.
Accordingly, the vectors of unknowns $\mm x_{\mathrm s}$ and $\mm x_{\mathrm a}$ correspond to the real Fourier coefficients of the generalized coordinates and the aerodynamic state variables, respectively.
The notation in \eref{rs} and \eref{ra} makes the coupled nature of the problem clear:
On the one hand, the aerodynamic forces, $\hat{\mm f}$ which depend on the aerodynamic state variables, $\mm x_{\mathrm a}$, enter the dynamic force balance in the structural domain.
On the other hand, the fluid mesh coordinates, $\hat{\mm\chi}$, which depend on the generalized coordinates, $\mm x_{\mathrm s}$, enter the governing equations in the fluid domain.
In the most general case, the parametrization constraint in \eref{rpc} is expressed as function of all (structure and fluid) unknowns.
Two variants have been developed to obtain the branch of the coupled solution, the coupled continuation of the solution path, and the coupled re-iteration of selected solution points.
The specific constraint equation depends on the variant, as explained later.
Before describing those variants, the general concept of the coupled solution is presented.
To understand the rationale behind the method development, it is crucial to emphasize that the computational effort for the fluid sub-problem is orders of magnitude higher than that associated with the structural sub-problem.

\subsection{Solution of the coupled problem via a fixed-point algorithm employing domain-specific solvers}
% IDEA
The idea is to solve the sub-problems in an alternating way, which can be interpreted as a fixed-point algorithm, as shown later.
A flow diagram of the coupling loop is given in \fref{sequentialcoupling}a.
To this end, a function $r_{\mathrm{pc}}^*(\mm x_{\mathrm s},\Omega)$ independent of $\mm x_{\mathrm a}$ is used instead of the more general function $r_{\mathrm{pc}}\left(\mm x_{\mathrm s},\mm x_{\mathrm a},\Omega\right)$ in \eref{rpc}.
A \emph{Newton-type solver} is then applied to obtain the solution of the augmented equation system
\east{
\mm R_{\mathrm s}\left(\mm X_{\mathrm s}\right) = \vector{
\mm r_{\mathrm s}\left(\mm x_{\mathrm s},\Omega,\hat{\mm f}^*\left(\mm x_{\mathrm s}\right)\right) \\
r_{\mathrm{pc}}^*(\mm x_{\mathrm s},\Omega)
} = \mm 0
%\quad \text{with respect to} \quad
\fk\quad
\mm X_{\mathrm s} = \vector{\mm x_{\mathrm s}\\ \Omega}
\fk \label{eq:extendedResidualS}
%\fp
}
with respect to the extended vector of unknowns $\mm X_{\mathrm s}$.
\\
% FLIN
In principle, one could adopt the aerodynamic forces, $\hat{\mm f}^{(\nu-1)}$, from the previous solution of the fluid sub-problem, where $\nu$ is the iteration number / counter of the coupling loop.
However, as shown in \cite{Berthold.2021,Berthold.2020}, much quicker convergence can be achieved when exploiting that the vibration-induced aerodynamic forces are in good approximation linear in $\mm x_{\mathrm s}$.
Thus, the expression
\east{
\hat{\mm f}^*\left(\mm x_{\mathrm s}\right) = \hat{\mm f}_{\mathrm{wake}} + \underbrace{\left(~\hat{\mm f}^{(\nu-1)}~-~\hat{\mm f}_{\mathrm{wake}}~\right)\cdot\frac{\mm e^{\mathrm T}\uhs}{\mm e^{\mathrm T}\uhs^{(\nu-1)}}}_{\hat{\mm f}_{\mathrm{vib}}} \label{eq:flin}
}
is used instead, where $\hat{\mm f}_{\mathrm{wake}}$ corresponds to the wake-induced aerodynamic forces present for rigid blades ($\uhs=\mm 0$), and $\mm e$ selects the fundamental Fourier coefficient of a representative generalized coordinate.
Good numerical performance was observed when selecting the coordinate of the resonant fixed-interface normal mode.
The construction of $\hat{\mm f}^*$ in \eref{flin} ensures that upon convergence, where $\mm x_{\mathrm s}^{(\nu)} = \mm x_{\mathrm s}^{(\nu-1)}$, we have consistently $\hat{\mm f}^*\left(\mm x_{\mathrm s}^{(\nu)}\right)=\hat{\mm f}^{(\nu-1)}$.
The wake-induced force $\hat{\mm f}_{\mathrm{wake}}$ generally depends on the rotor speed $\Omega$.
In the present study, good numerical performance was achieved already when this dependence was neglected, so that $\hat{\mm f}_{\mathrm{wake}}$ had to be computed only once for a nominal rotor speed.
\\
The problem dimension is $M(2\Hs+1)+1$, where $M=66$ in the numerical example and $\Hs$ is usually between 1 and 10.
Thus, the full Jacobian can be computed and factorized in each iteration without difficulties, which yields high convergence rates.
Analytical gradients are provided to increase efficiency.
% FLUID SUB-PROBLEM
The dimension of the fluid sub-problem is much higher, so that Newton-type solvers lead to prohibitive computation effort.
Instead, a \emph{pseudo-time solver} is applied to obtain the solution of \eref{ra} with respect to $\mm x_{\mathrm a}$ for the current estimate of $\mm X_{\mathrm s}^{(\nu)}$.
More specifically, a first order Euler backward scheme is applied.
This solution process is very similar to the classical solution method of a steady fluid problem.
In fact, the Jacobian of the steady solver is used during pseudo time integration which is an outcome of some simplifying assumptions \cite{Kersken2017}.
\\
% FIXED-POINT CHARACTER OF ALGORITHM
Applying the Newton-type solver to the above described structural sub-problem, using $\mm x_{\mathrm s}^{(\nu-1)}$ as initial guess, can be expressed in operator notation as $\mm X_{\mathrm s}^{(\nu)} = \mathcal S\left(\hat{\mm f}^{(\nu-1)},\mm x_{\mathrm s}^{(\nu-1)}\right)$.
Computing the mesh deformation, applying the pseudo-time solver to the fluid sub-problem, and computing the resulting aerodynamic force can be expressed in operator notation as $\hat{\mm f}^{(\nu)} = \mathcal F\left(\mm X_{\mathrm s}^{(\nu)}\right)$.
One iteration of the coupling loop corresponds to the chained application of these two operators,
\east{
\mm X_{\mathrm s}^{(\nu)} = \mathcal S \circ \mathcal F\left(\mm X_{\mathrm s}^{(\nu-1)}\right),
}
which corresponds to a \emph{fixed-point algorithm}.
The algorithm is terminated when the convergence condition $\|\mm X_{\mathrm s}^{(\nu)}-\mm X_{\mathrm s}^{(\nu-1)}\|<\varepsilon_{\mathrm{s}}$ is met.
It should be noted that relaxation strategies are commonly applied when solving problems of fluid-structure interaction using fixed-point algorithms.
When the fluid-to-structure mass ratio is low, as in the case of aero-elasticity of conventional gas and steam turbines, relaxation is not needed \cite{Degroote.2008} and excellent convergence was achieved in this study without relaxation.

\subsection{Coupled continuation of the solution path}
% TANGENT PREDICTOR AND ARC-LENGTH CONSTRAINT
In the present work, path continuation is implemented using a conventional predictor-corrector technique.
More specifically, a tangent predictor step is made,
\east{
\mm X_{\mathrm s} = \mm X_{\mathrm s}^{(n-1)} + \mm t\Delta s \fk
}
where $\mm X_{\mathrm s}^{(n-1)}$ is the previous point on the solution branch, $\mm t$ is the unit tangent at this point, and $\Delta s$ is the step length.
An arc-length constraint is imposed,
\east{
r_{\mathrm{pc}}^*\left(\mm x_{\mathrm s},\Omega\right) = \|\mm X_{\mathrm s} - \mm X_{\mathrm s}^{(n-1)}\|^2 - \Delta s^2 = 0\fk
}
where $\|\square\|$ denotes the Euclidean norm.
Tangent predictor and arc-length constraint are illustrated in \fref{pathcont}.
The predictor-corrector continuation loop starts at a certain $\Omega^{\mathrm{start}}$, typically in the linear regime, and ends at $\Omega^{\mathrm{end}}$.
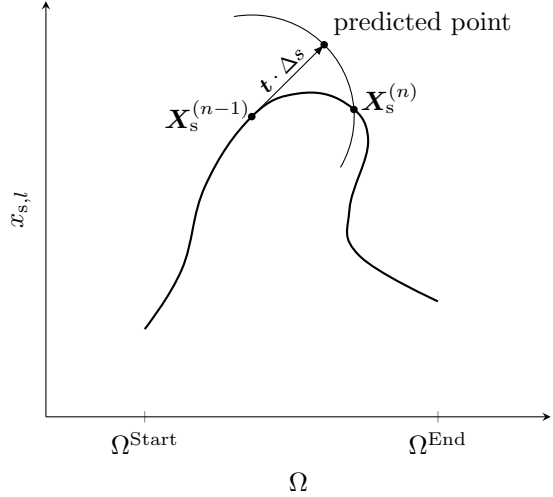
\begin{figure}
    \centering
        \begin{tikzpicture}[>=latex]
        \centering
        \begin{axis}[
            axis x line=bottom,  axis y line=left,
            width={0.5\linewidth},
            xlabel={$\Omega$},  ylabel={$ x_{\mathrm s,l}$},
            xmin= 0, xmax= 8,
            ymin= 0, ymax= 9,
            axis equal,
            ytick=\empty,
            xtick={0.7, 7},
            xticklabels={$\Omega^\mathrm{Start}$,$\Omega^\mathrm{End}$}]
            \addplot [thick, smooth, tension={0.7}, mark=none] table {
                0.7 1.9
                1.5 3.2
                2   5
                3   6.5
                4.0 7
                5   6.8
                5.5 6
                5.1 4.5
                5.3 3.5
                7   2.5
            };
            \draw [-latex](3,6.5) node [circle,fill,inner sep=1pt] {} node [above left, yshift=-9, xshift=3] {$\mm X_{\mathrm s}^{(n-1)}$} -- node [midway, above, yshift=0, xshift=2, rotate=45, ] {\footnotesize $ \mm t \cdot \Delta s$} ++(45:2.2) node [circle,fill,inner sep=1pt] {} node [above right] {predicted point} ;
            \draw (3,6.5) ++(-30:2.2) arc(-30:100:2.2);
            \draw (3,6.5) ++(4:2.2) node [circle,fill,inner sep=1pt] {} node [above right, yshift=-5, xshift=-1] {$\mm X_{\mathrm s}^{(n)}$} ;
        \end{axis}
        \end{tikzpicture}
    \caption{Numerical path continuation using a tangent predictor in combination with an arc-length constraint}
    \label{fig:pathcont}
\end{figure}
\\
% CORRECTOR
Starting from the prediction, the correction could be made, in principle, employing the coupled solver described in the previous subsection.
To robustly continue a strongly nonlinear, near-resonant solution branch, often several hundred solution points are needed.
Due to the large computational effort associated with the fluid solver, this may be practically infeasible.
A simple way to reduce the computation effort is to re-use the approximation $\hat{\mm f}^*$ in \eref{flin} for a number of points along the solution branch and only activate the coupling loop occasionally.
This algorithm is illustrated in \fref{sequentialcoupling}b.
\begin{figure}
\centering
\begin{subfigure}[t]{\textwidth}
\centering
\tikzstyle{decision} = [diamond, draw, text width=6em, text badly centered, node distance=4cm, inner sep=0pt]
\tikzstyle{block} = [rectangle, draw, text width=7em, text centered, rounded corners, minimum height=4em]
\tikzstyle{line} = [draw, -latex']
\tikzstyle{cloud} = [draw, ellipse,fill=red!20, node distance=3cm, minimum height=2em]

\begin{tikzpicture}[auto]

\node [block, text width=18em] (nlvib)
{{\Large\textbf{S}}tructure solver\\
update
$\hat{\mm f}^*\left(\mm x_{\mathrm s}\right)$
(if needed)
\\
solve
$
\left\{\begin{array}{l}
\mm r_{\mathrm s}\left(\mm x_{\mathrm s},\Omega,\hat{\mm f}^*\left(\mm x_{\mathrm s}\right)\right) = \mm 0 \\
r_{\mathrm{pc}}^*(\mm x_{\mathrm s},\Omega)  = \mm 0
\end{array}\right.
$
};

\node [block, below of=nlvib, node distance = 3cm, text width=18em] (trace)
{
{\Large\textbf{F}}luid solver\\
deform mesh $\hat{\mm\chi}^{(\nu)} = \hat{\mm\chi}\left(\mm x_{\mathrm s}^{(\nu)}\right) $\\
solve $\mm r_{\mathrm a}\left(\mm x_{\mathrm a},\Omega^{(\nu)},\hat{\mm\chi}^{(\nu)}\right) = \mm 0$\\
evaluate $\hat{\mm f}^{(\nu)} = \hat{\mm f}\left(\mm x_{\mathrm a}^{(\nu)}\right)$
};

\node [block, below of=trace, node distance = 2.5cm, text width=12em] (init)
{
initialize coupling loop\\
$\nu \leftarrow 0$,
$\mm X_{\mathrm s}^{(\nu)} \leftarrow \mm X_{\mathrm s}^{(n)}$
};

\node [left of=trace, node distance = 5cm] (anchorconv){};

\node[decision, above of=anchorconv, node distance = 1.5cm, aspect=1.3](convergencedecision){converged ?};

\path [line] (trace) -- ++(4.5,0) node(anchorstore){}  |- node [near start, left] {\small $\nu \leftarrow \nu + 1$} node[pos=0.6, above]{$\hat{\mm f}^{(\nu-1)}$} (nlvib);
\path [draw, dotted] (anchorstore) -- ++(1cm,0);
\path [line] (anchorconv.center) -- (trace);
\path [line] (init) -- (trace);
\path [draw] (convergencedecision) -- node[midway]{no} node[at end, below]{$\mm x_{\mathrm s}^{(\nu)},\Omega^{(\nu)}$}(anchorconv.center);
\path [draw] (convergencedecision.west) -- node[midway]{yes} ++(-1cm,0) node(anchordotted){};
\path [draw, dotted] (anchordotted) -- ++(-1cm,0);
\path [line] (nlvib) -| (convergencedecision);

\path [draw, dotted] (init.west) ++(-2cm, 0) -- ++(1cm,0) node(anchorinitall){};
\path [line] (anchorinitall.center) -- (init);

\end{tikzpicture}
\caption{}
\label{fig:sequentialcoupling_a}
\end{subfigure}

\vspace{0.5cm}
\centering
\begin{subfigure}[t]{\textwidth}
\centering
\tikzstyle{decision} = [diamond, draw, text width=4em, text badly centered, node distance=4cm, inner sep=0pt]
\tikzstyle{block} = [rectangle, draw, text width=7em, text centered, rounded corners, minimum height=4em]
\tikzstyle{line} = [draw, -latex']
\tikzstyle{cloud} = [draw, ellipse,fill=red!20, node distance=3cm, minimum height=2em]

\begin{tikzpicture}[auto]

%STRUCTURE SOLVER
\node [block] (nlvib)
{{\Large \textbf{S}}truct. solver};

%FLUID SOLVER
\node [block, below of=nlvib, node distance = 3cm] (trace)
{{\Large \textbf{F}}luid solver};

%PREDICTOR STEP
\node [block, above of=nlvib, node distance = 3.5cm, text width=9em] (predstep)
{predictor step\\$n \leftarrow n + 1$\\$\mm X_{\mathrm s} = \mm X_{\mathrm s}^{(n-1)} + \mm t\Delta s$};

%SWITCH MEMORY
\draw node[cute spdt up, right of=nlvib, node distance = 2.5cm] (mem) {}
(mem.out 1) node[above] {1}
(mem.out 2) node[below] {2};
\path [line] (mem.in) -- (nlvib);
\path [line] (trace) -- ++(3.5,0) node(branchnode){} |- node[left, pos=0.3]{$\nu \leftarrow \nu + 1$} (mem.out 2);

%MEMORY BLOCK
\node [block, right of=trace, node distance = 5cm, text width=3em,  minimum height=2em] (memblock)
{store $\hat{\mm f}^{(\nu)}$\\ \Large $\circlearrowleft$};
\path [line] (branchnode.center) -- (memblock);
\path [line] (memblock) |- (mem.out 1);

%SWITCH TO PREDICTOR STEP
\draw node[cute spdt up, left of=nlvib, node distance = 2.5cm, xscale=-1] (pred) {}
(pred.out 1) node[above] {1}
(pred.out 2) node[below] {2};
\draw (nlvib) -- (pred.in);
\draw node[left of=pred-out 1, node distance = 0.7cm] (anchor1) {};
\draw node[left of=pred-out 2, node distance = 0.7cm] (anchor2) {};
%\path [line] (pred.out 2) -- ++(-1,0) |- (trace);

%SWITCH FROM PREDICTOR STEP
\draw node[cute spdt down, below of=predstep, yshift=-0.4cm, rotate=-60, xshift=0.4cm] (switchpred){}
(switchpred.out 1) node[right] {2}
(switchpred.out 2) node[left] {1};
\draw (predstep) -- (switchpred.in);
\path [line] (switchpred.out 2) -- (nlvib);

%INITIALIZATION BLOCKS
\node [block, below of=trace, node distance = 1.5cm, text width=10em, minimum height=2em] (initcoupling)
{initialize coupling loop};
\node [block, left of=initcoupling, node distance = 5cm, text width=12em] (initialization)
{initialization\\$n \leftarrow 0$,\\ $\mm X_{\mathrm s}^{(n)} \leftarrow (\mm x_{\mathrm s}^\mathrm{Start}, \Omega^\mathrm{Start})$,\\switches: pos. 2};
\path [line] (initialization) -- (initcoupling);
\path [line] (initcoupling) -- (trace);

%TRIGGER COUPLING LOOP DECISION
\node [decision, above of=anchor1, node distance = 1.5cm, aspect=1.5] (triggercoupling)
{\small trigger coupling loop?};
\path [line] (pred.out 1) -| (triggercoupling);
\path [line] (triggercoupling) |- node[pos=0.5, above]{no} (predstep);

%SWITCH COUPLING LOOP BLOCK
\node [block, left of=triggercoupling, node distance = 4cm, text width=8em] (switchoncoupling)
{switches: pos. 2, \\ go to ``initialize coupling loop''};
\path [line] (triggercoupling) --  node[midway, above]{yes} (switchoncoupling);

%TRIGGER CONTINUATION DECISION
\node [decision, below of=anchor2, node distance = 1.3cm, aspect=1.6, text width=7em] (triggercontinuation)
{\tiny $\|\mm X_{\mathrm s}^{(\nu)}-\mm X_{\mathrm s}^{(\nu-1)}\|<\varepsilon_{\mathrm{s}}$};
\path [line] (pred.out 2) -| (triggercontinuation);
\path [line] (triggercontinuation) |- node[midway, below]{no} (trace);

%SWITCH COUPLING LOOP BLOCK
\node [block, left of=triggercontinuation, node distance = 4.1cm, text width=9em] (switchoncontinuation)
{switches: pos. 1,\\$\mm X_{\mathrm s}^{(n)} \leftarrow \mm X_{\mathrm s}^{(\nu)}$,\\ go to predictor step};
\path [line] (triggercontinuation) --  node[midway, below]{yes} (switchoncontinuation);

%\path[line] (predswitch.east) to[-latex] (nlvib);
%\node [right of=predswitch, xshift=-0.5cm] {1};
%\node [left of=predswitch, xshift=0.5cm] {2};

%\path [line] (nlvib) -- node [midway] {$\mm x_{\mathrm s}^{(\nu)},\Omega^{(\nu)}$} (trace);
%\path [line] (trace) -- ++(5,0)  |- node [near start, right] {$\nu \leftarrow \nu + 1$} (nlvib);

\end{tikzpicture}
\caption{}
\label{fig:sequentialcoupling_b}
\end{subfigure}
\caption{Flow diagram of (a) the coupling loop, (b) the coupled continuation of the solution path}
\label{fig:sequentialcoupling}
\end{figure}
\\
% ACTIVATION CRITERIA FOR THE COUPLING LOOP
Different criteria are conceivable for activating the coupling loop, which may account, for instance, for a change of amplitude or rotor speed.
In the present work, the coupling loop was simply activated after a fixed number of solution points.
As alternative, the coupling loop was activated when the vibrational deflection shape had changed to a certain extent.
To quantify the change of the deflection shape, the correlation measure
\east{
\beta\left(\mm x_{\mathrm s}\right) = 1-\left(\frac{\mm x_{\mathrm s}^{\mathrm T} \mm x_{\mathrm s}^{(\nu-1)}}{\|\mm x_{\mathrm s}\| \|\mm x_{\mathrm s}^{(\nu-1)}\|}\right)^2 \label{eq:MAC}
}
was used.
Upon completion of the coupling loop, the approximation of the aerodynamic force, $\hat{\mm f}^*$ in \eref{flin}, is updated and used to obtain the subsequent solution points (for which the coupling loop is inactive).

\subsection{Coupled re-iteration of selected solution points}
An important weakness of the coupled continuation is that it is inherently sequential; the computationally involved fluid solver iterations cannot be run in parallel.
This motivated the development of the coupled re-iteration approach.
First, the whole solution branch is computed in a decoupled way, \ie, using the structural solver only.
For the aerodynamic force, a low-fidelity state-of-the-art model can be used; \eg the linear superposition of the wake-induced loading and an approximation of the vibration-induced loading using a single modal aerodynamic influence coefficient.
Second, a relevant subset of solution points along the branch are selected.
Here, regions that are of high engineering relevance, such as the resonance peak, can be resolved with a finer spacing of points.
Finally, starting from this subset of points of the decoupled solution, the coupled solver is applied to obtain a corresponding point on the branch of the coupled solution.
This idea is illustrated in \fref{ParallelContinuation}.
It should be emphasized that the coupled solver can be applied to all selected solution points in parallel.
\fig{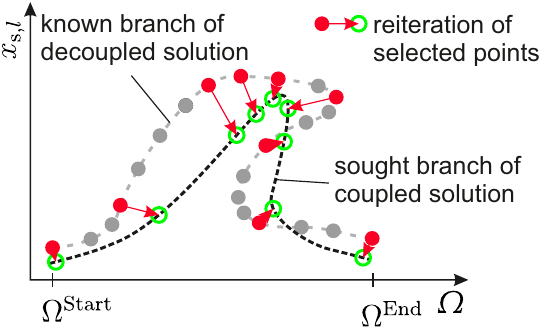}{Coupled re-iteration of selected solution points
}{0.65}
\\
% PARAMETRIZATION IN THE CASE OF RE-ITERATION
To define where the corresponding point ends up on the new solution branch, it is required that the new solution $\mm X_{\mathrm s}$ lies on the hyper-plane orthogonal to the tangent $\mm t$ at the point $\mm X_{\mathrm s}^{\mathrm{ref}}$ on the old solution branch,
\east{
r_{\mathrm{pc}}^*\left(\mm X_{\mathrm s}\right) = \mm t^{\mathrm T}\left(\mm X_{\mathrm s} - \mm X_{\mathrm s}^{\mathrm{ref}}\right) = 0\fp
}
It is expected that this parametrization constraint is relatively robust to a substantial change in topology between old and new solution branch, compared to, for instance, the local parametrization obtained by imposing a fixed rotor speed $\Omega$.
\section{Numerical Results}
The forced response near the resonance of the lowest-frequency mode family with engine order 20 is analyzed in this section.
The depicted results were obtained for truncation orders $\Hs=1$ and $\Ha=2$.
This way, the effects of multi-harmonic fluid-structure interaction are suppressed, much like in the case of the influence-coefficients-based state-of-the-art methods.
This permits a clearer interpretation of the deviations between those methods and the proposed fully coupled methods:
The deviations can only be due the amplitude dependence of the aerodynamic forces and the nonlinear superposition of wake- and vibration-induced flow.
It should be remarked that simulations with higher truncation orders have been carried out.
The results show only slight quantitative deviations and are not depicted for brevity.
In the following, first, the results of the coupled analysis are compared against the state of the art.
Then the numerical performance of the distinct variants of the proposed method is assessed.

\subsection{Comparison of coupled analysis results against the state of the art}
% DESCRIPTION OF STATE OF THE ART
It is the current state of the art to model the aerodynamic forces in a linear way using influence coefficients.
This corresponds to solving \eref{extendedResidualS} using as approximation for $\hat{\mm f}^*$:
\east{
\hat{\mm f}^*_1 = \hat{\mm f}_{\mathrm{wake},1} + \mm G_{\mathrm a} \hat{\mm u}_1^{\mathrm s} \quad \hat{\mm f}_h^* = \mm 0 ~~ h=0,2,\ldots,\Hs\fp
}
Herein, $\mm G_{\mathrm a}$ is the aerodynamic influence coefficient matrix, formulated in the space spanned by the component modes.
Note that the approximation is inherently limited to the fundamental harmonic.
Two variants are distinguished, the single-aerodynamic-influence-coefficient (\emph{single AIC}) method, and the full-matrix-of-influence-coefficients (\emph{AIC matrix}) method.
\\
% FULL MATRIX
For the AIC matrix method, $M$ CFD simulations are carried out, one for each component mode, where the structure vibrates with unit amplitude in the respective component mode.
The resulting fundamental Fourier coefficient of the generalized aerodynamic force is used as respective column of the matrix $\mm G_{\mathrm a}$.
\\
% SINGLE COEFFICIENT
For the single AIC method, only a single CFD simulation is carried out, where the structure vibrates with unit amplitude in the resonant normal mode, $\mm\varphi$, of the linearized structure (sticking contact conditions).
By projecting the resulting fundamental harmonic of the generalized aerodynamic force onto $\mm\varphi$, one obtains a scalar coefficient $\alpha$.
A consistent matrix $\mm G_{\mathrm a}^\alpha$ is then formulated,
\east{
\mm G_{\mathrm a}^\alpha = \left(\mm M\mm\varphi\right)\,\alpha\,\left(\mm M\mm\varphi\right)^{\mathrm H}\fk 
}
where $\mm M$ is the mass matrix, and it is assumed that $\mm\varphi$ is mass-normalized ($\mm\varphi^{\mathrm H}\mm M\mm\varphi = 1$).
\fig[htb]{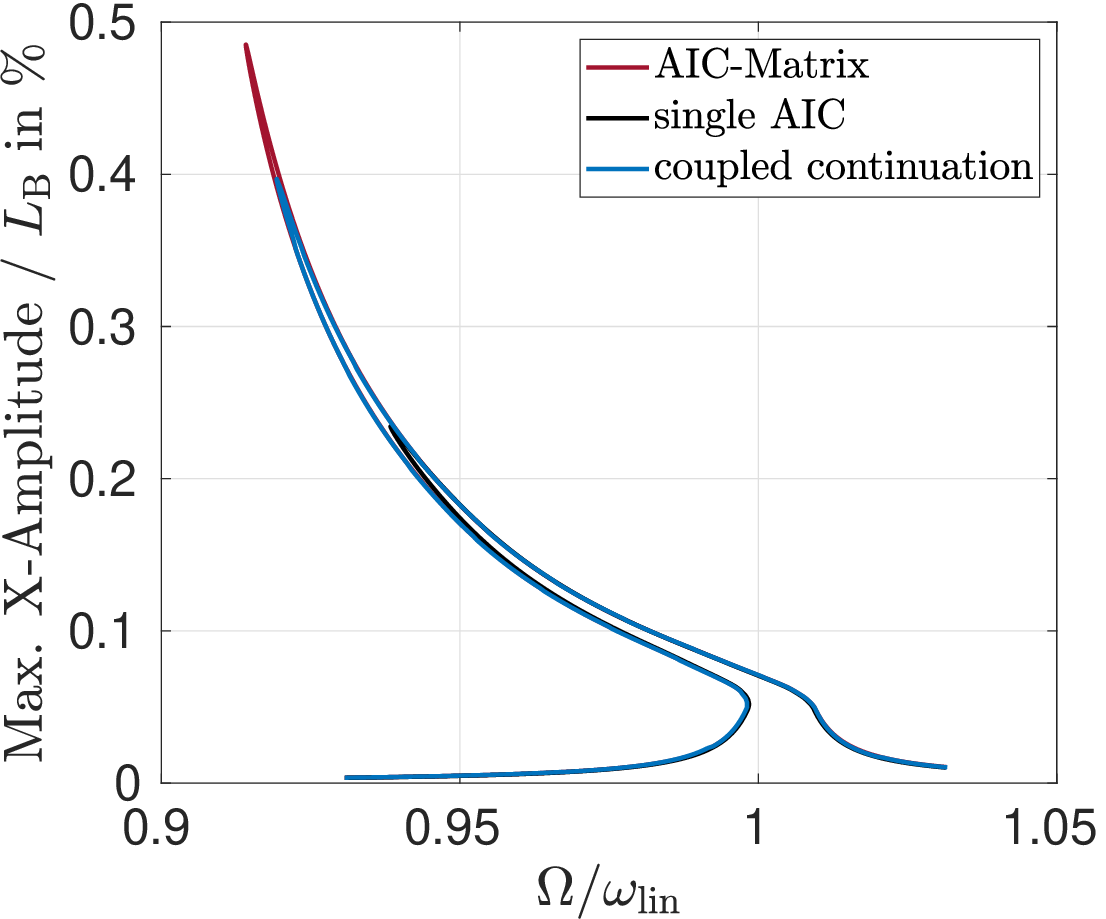}{
Amplitude-frequency curve: Coupled continuation vs. AIC methods
}{0.65}%
\\
% FREQUENCY RESPONSE RESULTS
The forced response is illustrated in \fref{FRF_FIC_1_SIC_1_Iter_0_Orig_0_SeqMAC_0_SeqInt_1_FICnew_0}.
As amplitude measure, the maximum $X$-displacement of the sensor node (see \fref{T16001}) is used, normalized by the blade length $L_{\mathrm B}$.
The abscissa shows the rotor speed normalized by the resonance speed for the linearized structure (sticking contacts).
All methods predict a sharp resonance peak (indicating very light damping) and a softening characteristic (resonance frequency decreasing with amplitude.
These aspects are typical for pronounced normal contact interactions, in the form of temporary opening of (parts of) the interlocked shroud joint, without much sliding in the tangential contact plane (which would lead to a substantial increase of frictional damping with vibration level).
This is confirmed by the contact behavior illustrated in \fref{contactBehavior}, which indicates a rolling-type motion at intermediate and high amplitudes.
\begin{figure}[htb]
\centering
\begin{subfigure}[t]{0.29\textwidth}
 \centering
 \includegraphics[width=\textwidth]{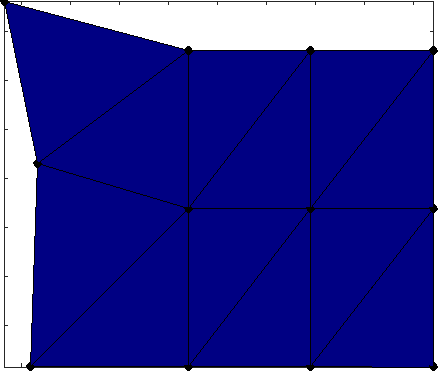}
 \caption{}
 \label{fig:cstatePntC}
\end{subfigure}
\begin{subfigure}[t]{0.29\textwidth}
 \centering
 \includegraphics[width=\textwidth]{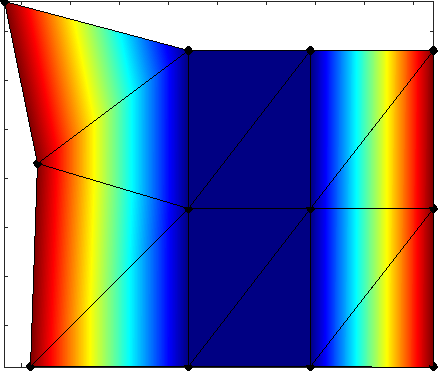}
 \caption{}
 \label{fig:cstatePntB}
\end{subfigure}
\begin{subfigure}[t]{0.29\textwidth}
 \centering
 \includegraphics[width=\textwidth]{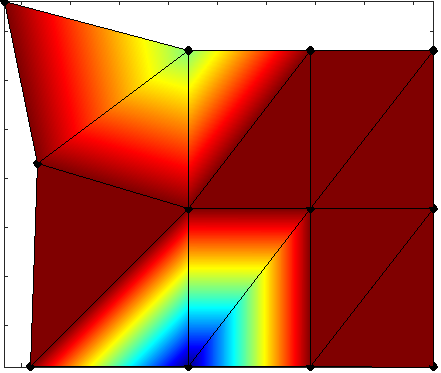}
 \caption{}
 \label{fig:cstatePntA}
\end{subfigure}
\caption{Contact behavior at three representative point along the upper solution branch of the forced response in \fref{FRF_FIC_1_SIC_1_Iter_0_Orig_0_SeqMAC_0_SeqInt_1_FICnew_0} (coupled continuation result): (a) low amplitude at $\Omega/\omega_{\mathrm{lin}} = 1.03$; (b) intermediate amplitude at $\Omega/\omega_{\mathrm{lin}} = 0.98$; (c) high ampliude at $\Omega/\omega_{\mathrm{lin}} = 0.93$. Color blue represents permanently sticking contact elements. Red indicates elements, which separate at least once during the vibration cycle. Intermediate colors indicate amount of sliding during the vibration cycle.}
\label{fig:contactBehavior}
\end{figure}
The combination of softening with light damping yields turning points with respect to the excitation frequency / rotor speed, a phenomenon which is probably best known from the Duffing oscillator.
The overhanging branch between the two turning points is known to be unstable.
\\
% DISCUSSION OF DISCREPANCY
While the results obtained by the three methods agree qualitatively, substantial quantitative differences appear.
All results align well at sufficiently low amplitudes, where the shroud contact is permanently sticking (\fref{contactBehavior}a) and the system behaves in good approximation linear.
In that case, the vibrational deflection shape is dominated by the resonant linear mode, $\mm\varphi$, and the single AIC method is consistent with the AIC matrix method.
At higher amplitudes, dynamic contact interactions occur, including sliding and temporary liftoff phases during the vibration cycle.
Thus, the effective deflection shape changes, and the single AIC method is no longer consistent with the AIC matrix method.
Remarkably, the results differ with respect to the resonant amplitude by a factor of two.
In this numerical example, apparently, the deflection shape changes in such a way that the effective aerodynamic damping is lower.
The aerodynamic damping for the linear mode (sticking contacts) about $1\%$ logarithmic decrement.
A strong sensitivity of the aerodynamic damping on the deflection shape was also found in the flutter analyses in \cite{Berthold.2020,Berthold.2021}.
The coupled continuation result is somewhere between those of the two AIC methods in this case.
Like the coupled continuation method, the AIC matrix method is generally able to account for a change of the deflection shape.
The difference between AIC matrix method and coupled continuation could be due to the amplitude-dependence of the aerodynamic force or due to the invalid superposition of wake- and vibration-induced flow, and this is further analyzed in the following.
\fig[htb]{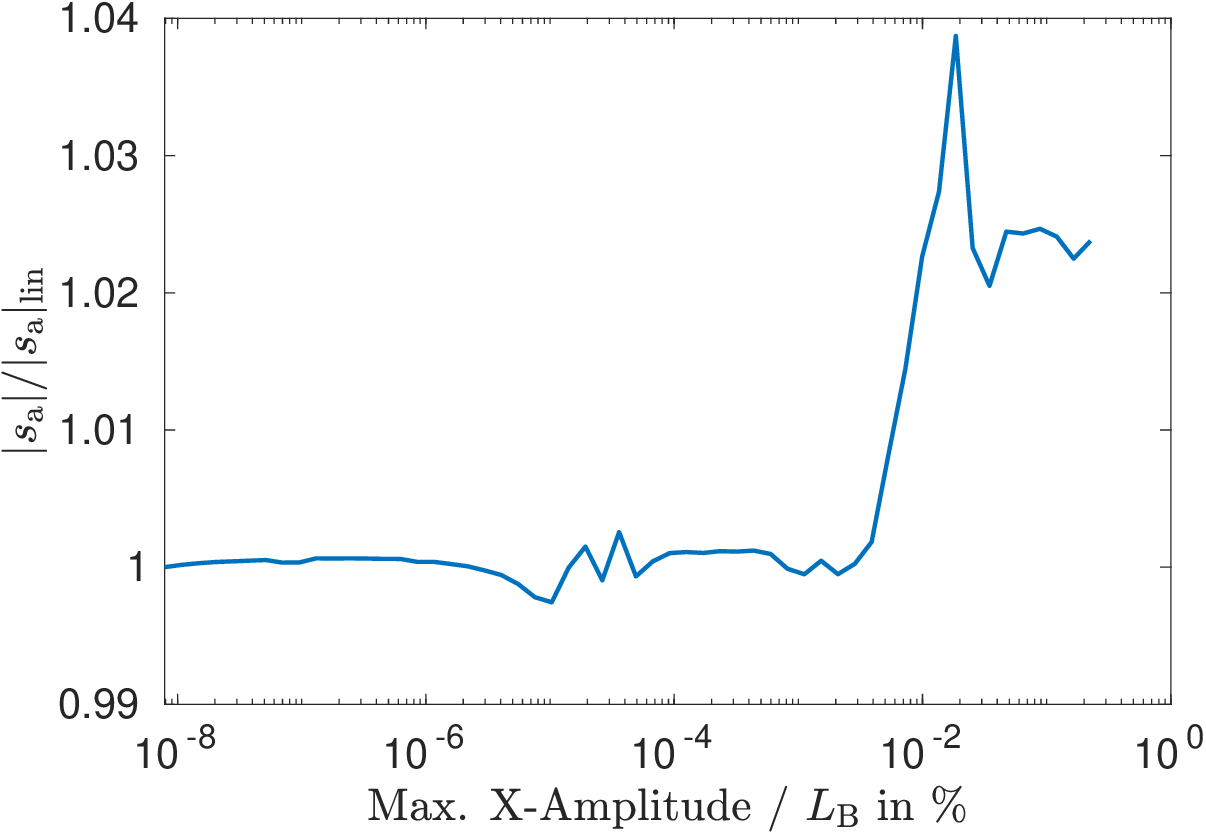}{
Modal aerodynamic stiffness (according to \eref{sa}) vs. amplitude
}{0.65}
\fig[htb]{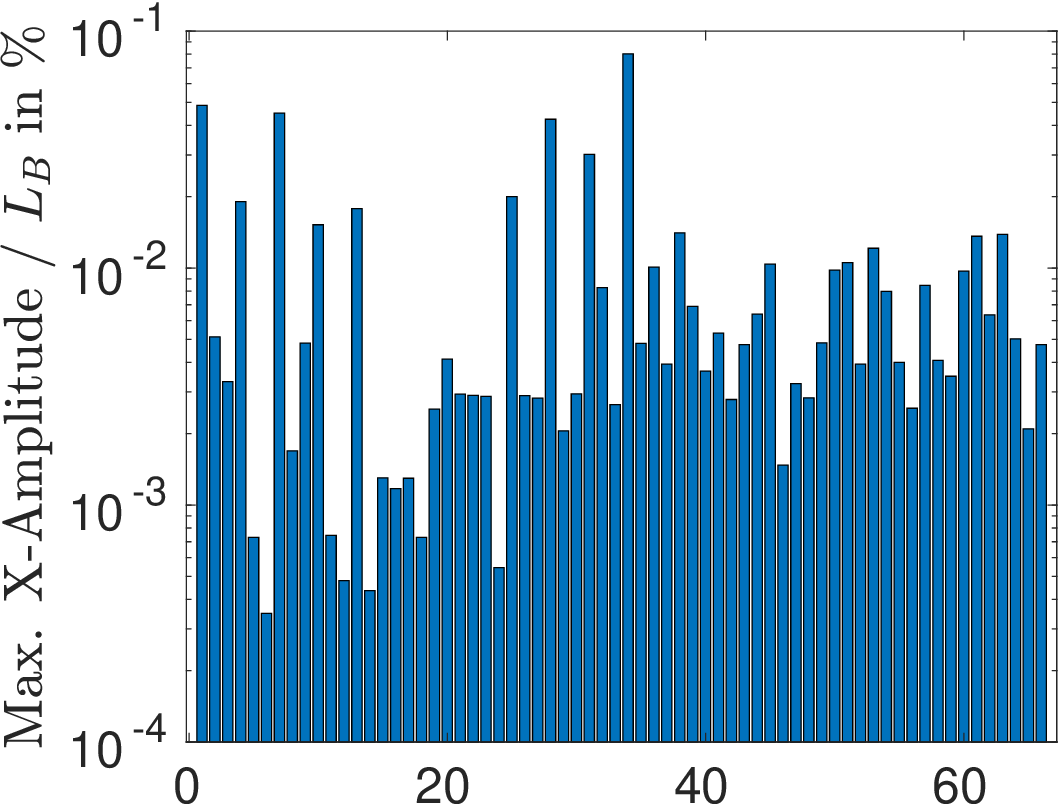}{Amplitude of X-Displacement at sensor node for component modes with unit amplitude}{0.65}
\\
% AMPLITUDE-DEPENDENCE HYPOTHESIS
To analyze the amplitude-dependence of the aerodynamic force, a ratio $s_\mathrm{a}$ is analyzed, which is defined as
\east{
s_{\mathrm a} = \frac{{\mm e}^{\mathrm T} \hat{\mm f}}{{\mm e}^{\mathrm T}\uhs}\fp \label{eq:sa} 
}
% aerodynamic stiffness is usually defined as kappa=Im{aeroforce}/q*... and thus is propably not suited to denote s_a
Herein, ${\mm e}$ selects the entry corresponding to the fundamental Fourier coefficient associated with the resonant fixed-interface normal mode (also used in \eref{flin}).
$\hat{\mm f}$ is obtained by a CFD simulation where the blade vibrates in this mode.
The amplitude of this mode is varied and the results are shown in \fref{AmpVariationForceDOF37}.
In the plot, $\left|s_{\mathrm a}\right|$ is normalized to have a value of 1 in the linear case.
For amplitudes up to $2\cdot 10^{-3}\% L_{\mathrm B}$, the deviation from linearity is in the numerical noise floor.
Then a distinct but small deviation of up to $3.8\%$ is visible, which is induced by a slight change of the shock dynamics in the tip region of the blade.
This small deviation alone does not seem to have the potential to explain the discrepancy between AIC matrix method and coupled continuation observed in \fref{FRF_FIC_1_SIC_1_Iter_0_Orig_0_SeqMAC_0_SeqInt_1_FICnew_0}.
It should be noted that the influence coefficients are obtained for finite amplitudes.
More specifically, a unit amplitude is prescribed.
To get an idea of the order of magnitude of this amplitude, it is depicted in \fref{sensorNodeDispOfBasisVectors} for the X-displacement of the sensor node.
Note that some amplitudes are very low because the deflection shape is mainly oriented orthogonal to the X-direction.
Clearly, the amplitudes are small and largely negligible when compared to the forced response amplitudes in \fref{FRF_FIC_1_SIC_1_Iter_0_Orig_0_SeqMAC_0_SeqInt_1_FICnew_0}.
It should be remarked that the evaluation with finite amplitudes (and with harmonic balance) was chosen over a linear CFD solver in order to achieve a consistent turbulence modeling, which requires second-harmonic terms that cannot be captured by a linear solver.
\fig[htb]{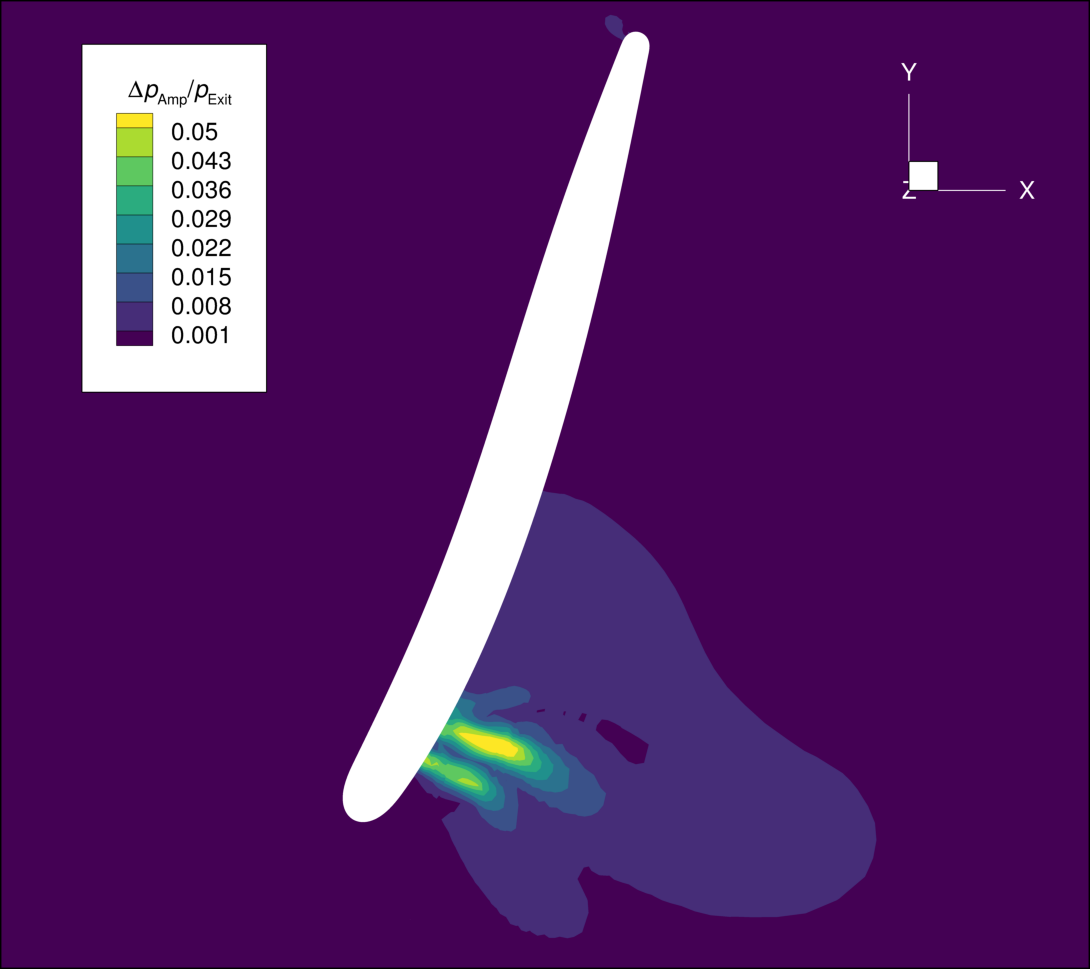}{Deviation $\Delta\hat p_1$ between the pressure field obtained by the coupled solver and the linear superposition of wake- and vibration-induced pressure field, normalized by the exit pressure, depicted at \SI{80}{\percent} blade height.}{0.75}
\\
% INVALID SUPERPOSITION HYPOTHESIS
Next, it is analyzed to what extent the superposition of wake- and vibration-induced flow holds.
To this end, the coupled solution point with the highest vibration amplitude is considered.
The pressure field induced solely by the vibration, $\hat p_{\mathrm{vib}}$, is computed, and the deviation $\Delta \hat p$,
\east{
\Delta \hat p = \hat p - \left(\hat p_{\mathrm{wake}} + \hat p_{\mathrm{vib}}\right)\fk
}
between the sum of wake- and vibration-induced pressure field, on the one hand, and the pressure field $\hat p$ obtained from the coupled solver, on the other hand, is determined.
The fundamental Fourier coefficient, $\Delta \hat p_1$, is depicted in \fref{phat80} at \SI{80}{\percent} blade height, normalized by the exit pressure.
Apparently, the superposition fails, in particular, near the location of the shock (see also \fref{mach}), where the superposition error amounts to $5.5\%$.

\subsection{Assessment of the numerical performance of the developed methods}
\begin{figure}[htb]
\centering
 \begin{subfigure}[t]{0.49\textwidth}
 \centering
 \input{figs/frfseqzoom.tex}
 \caption{}
\end{subfigure}
\hfill
\begin{subfigure}[t]{0.49\textwidth}
 \centering
 \includegraphics[width=\textwidth]{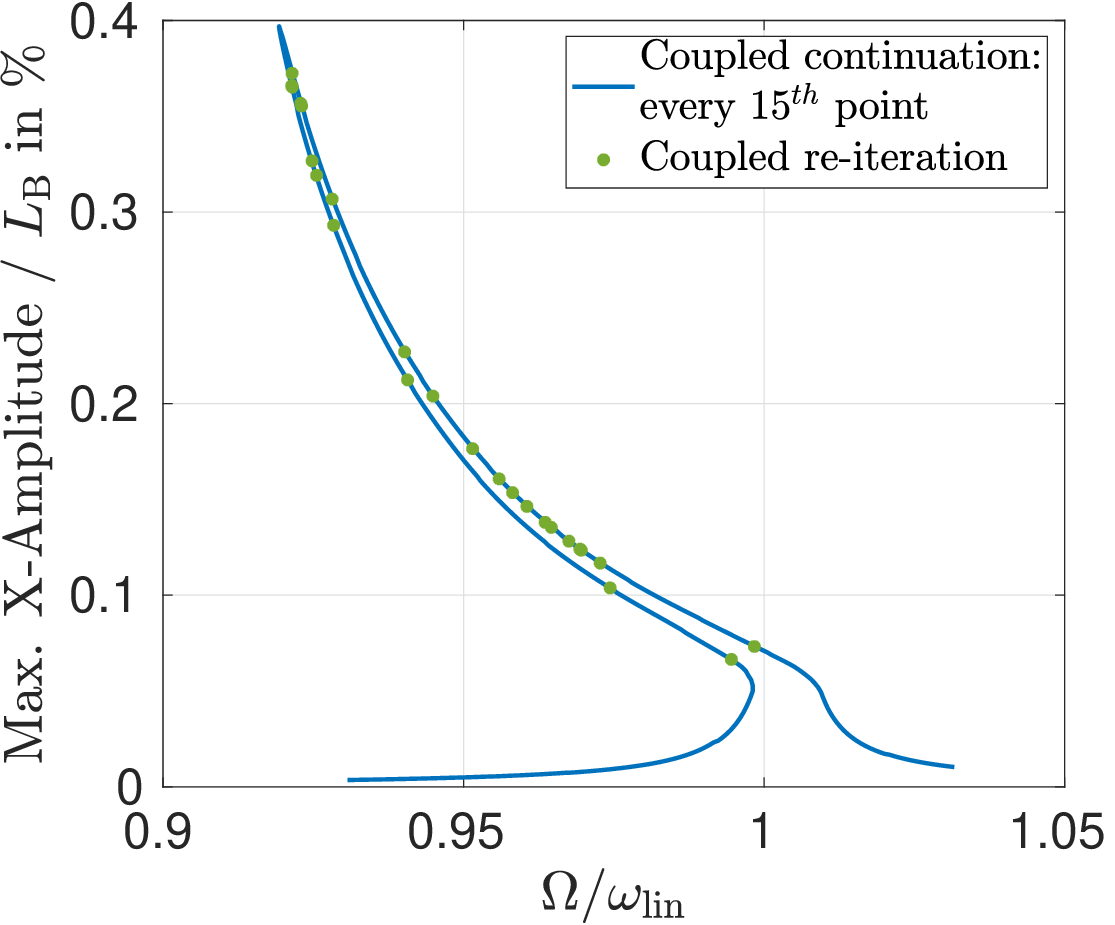}
 \caption{}
\end{subfigure}
\caption{
Amplitude-frequency curve:
(a) coupled continuation with two different activation criteria for the coupling loop;
(b) coupled continuation vs. coupled re-iteration
\label{fig:coupledMethodsOne}
}
\end{figure}
\begin{figure}
\centering
\begin{subfigure}[t]{0.49\textwidth}
 \centering
 \includegraphics[width=\textwidth]{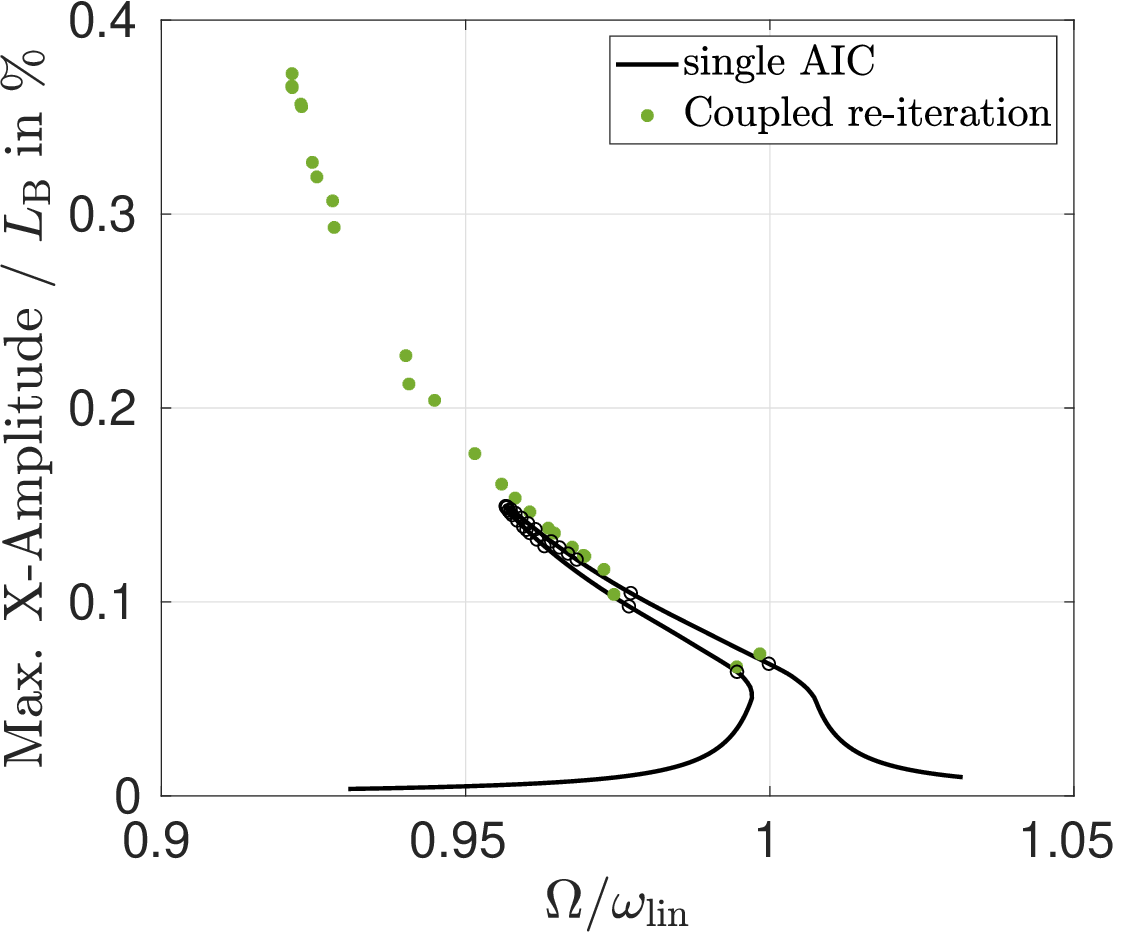}
 \caption{}
\end{subfigure}
\begin{subfigure}[t]{0.49\textwidth}
 \centering
 \includegraphics[width=\textwidth]{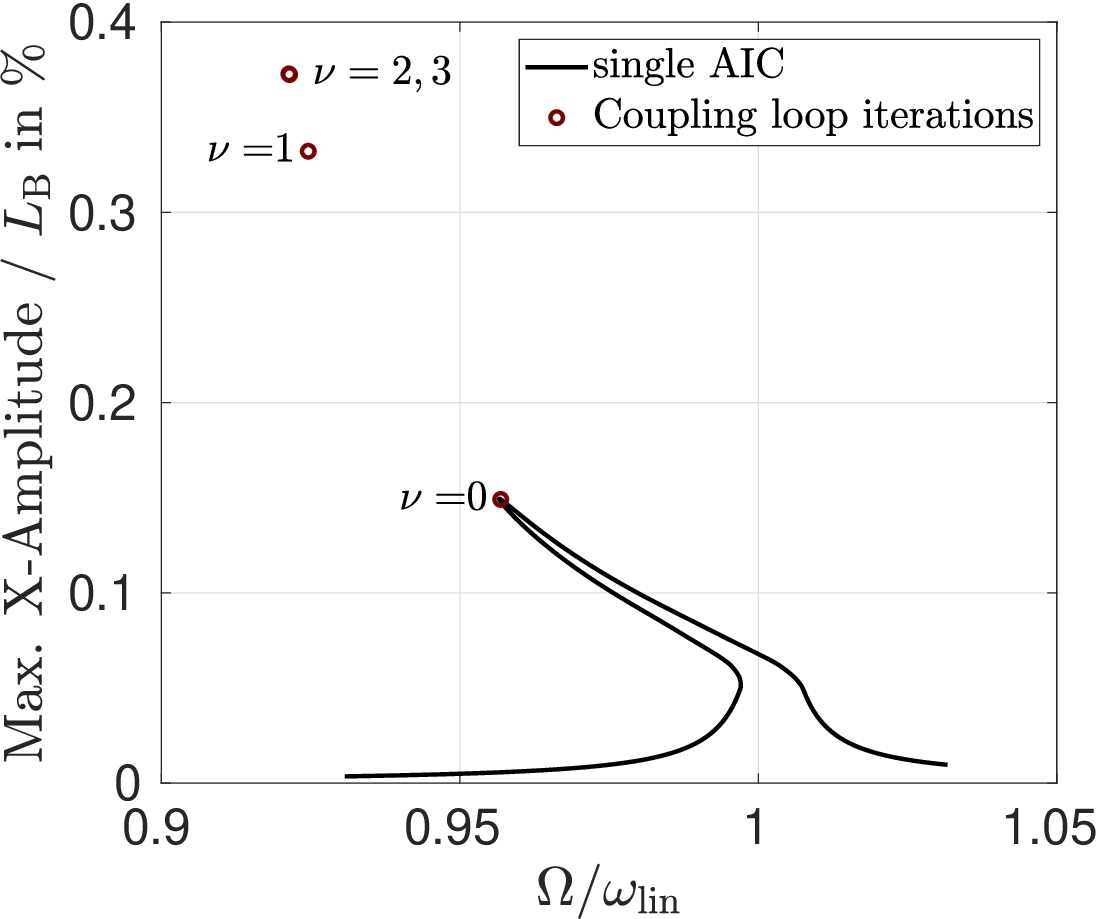}
 \caption{}
\end{subfigure}
\caption{
Amplitude-frequency curve obtained by coupled re-iteration:
(a) initial points (black circles) on old branch and final points (green dots) on new branch;
(b) sequence of points generated by coupling loop iterations for a representative point near the resonance peak
\label{fig:coupledMethodsTwo}
}
\end{figure}
\begin{table}[ht]
\centering
\begin{tabular}{llll}
Method & wct (fluid+solid) /\SI{}{\hour} & tch (fluid+solid) /\SI{}{\hour} \\ \toprule
steady CFD & 3.4+0=3.4 & 217+0=217 \\
single AIC & 9.45+0.19=9.64 & 605+0.19=605.19 \\
AIC matrix & 10.2+0.28=10.48 & 43084.8+0.38=43085.18 \\
coupled cont. (coupling every $15^{\mathrm{th}}$ point) & 290.0+0.30=290.3 & 18560+0.3=18560.3\\
coupled cont. (cpl. if mode shape changed) & 144.0+0.30=144.3 &  9216.0+0.3=9216.3\\
coupled re-iteration (parallel) & 29.3+0.02=29.32 & 32286.8+0.02=32286.82\\
\end{tabular}
\caption{Comparison of wall clock time (wct) and total core hours (tch) for the different methods.}
\label{tab:computationalcosts}
\end{table}
%
% COMPARISON OF DIFFERENT ACTIVATION CRITERIA
For the coupled continuation results shown so far, the coupling loop was simply activated every $15^{\mathrm{th}}$ point.
As alternative, the coupling loop is activated if the vibrational deflection shape (or mode shape) changed, in accordance with the correlation measure in \eref{MAC} and with the threshold $\beta>0.02$, using the mode shape of the previous coupled solution point as reference.
The results are compared in \fref{coupledMethodsOne}a.
The amplitude-frequency curves are largely indistinguishable.
Minor deviations occur only at low amplitudes before the coupling loop is activated for the second time, see zoom in \fref{coupledMethodsOne}a.
With the mode-shape-based activation criterion, the number of coupled iterations along the continued branch is reduced from $29$ to $16$.
This reduced the wall clock time of two weeks by two days (\tref{computationalcosts}).
There is certainly potential for more sophisticated and problem-adapted criteria for activating the coupling loop; but these are considered beyond the scope of the present work.
\\
% COUPLED CONTINUATION VS. COUPLED RE-ITERATION
Coupled continuation and coupled re-iteration results are compared in \fref{coupledMethodsOne}b.
The green dots correspond to the re-iterated solution points and are in excellent agreement with the solution branch obtained by coupled continuation.
A total of 25 points is selected for re-iteration.
These are indicated as black circles on the old solution branch (black solid line) in \fref{coupledMethodsTwo}a, whereas the re-iterated points are indicated as green dots.
Comparing closely \fref{FRF_FIC_1_SIC_1_Iter_0_Orig_0_SeqMAC_0_SeqInt_1_FICnew_0} and \fref{coupledMethodsTwo}a, one may notice that the single AIC solution branch differs.
This is because not the consistent influence coefficient ($\alpha$) obtained by CFD was used but a generic value in the same order of magnitude ($1.5\%$ instead of $1\%$ logarithmic decrement).
Still, consistent results of both coupled analysis methods are achieved.
\fig{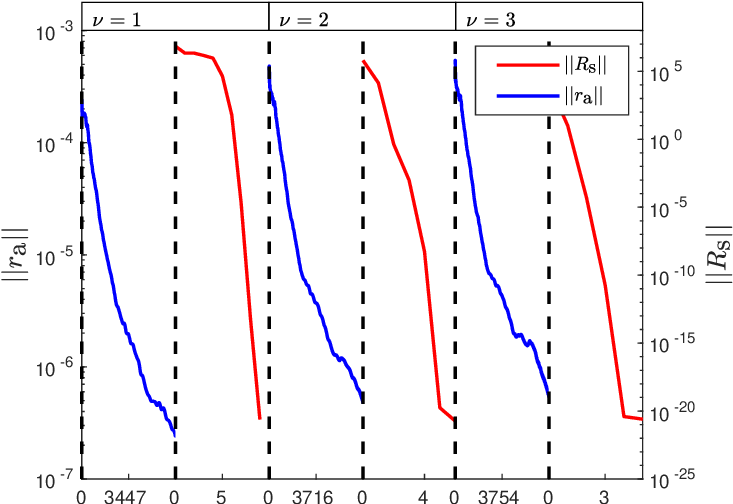}{
Decay of residuals in fluid and structural domain for the coupling loop iterations illustrated in \fref{coupledMethodsTwo}b.
}{0.75}
\\
% CONVERGENCE BEHAVIOR
For a representative initial point near the resonance peak, the sequence of points generated during the coupled re-iterations is shown in \fref{coupledMethodsTwo}b.
Recalling that the iterated solution points are required to lie in the hyper-plane orthogonal to the tangent at the selected point on the old solution branch, one may wonder why the iterated points are not on a straight line in \fref{coupledMethodsTwo}b.
The reason for this is that the orthogonality holds only in the space of the unknowns, $\mm X_{\mathrm s}$, not in the amplitude-frequency space depicted here, since the amplitude is a nonlinear function (weighted Euclidean norm) of $\mm X_{\mathrm{s}}$.
Further, it is remarkable that only three iterations are needed.
The corresponding decay of the residual norms in the fluid and the structure domain is depicted in \fref{convergenceIterationsCoupledSolver}.
That result is in line with the results of the coupled flutter analysis in \cite{Berthold.2020,Berthold.2021}.
However, a very good initial guess was available in those studies, whereas the initial solution point deviates substantially (at least in terms of amplitude) in the present forced response example.
The fact that still three iterations are sufficient, and the residual norms decay well, demonstrates the high robustness of the proposed numerical methods.
As can be seen in \fref{convergenceIterationsCoupledSolver}, the Newton-type structural solver requires around 5 iterations, whereas the pseudo-time fluid solver requires a few thousands of iterations.
This is expected, because of the generally higher convergence rate of Newton-type solvers and much smaller number of unknowns in the structural domain.
\\
% COMPUTATIONAL EFFORT
In general, one may expect that the coupling loop converges more rapidly during the continuation, because the initial guess then always remains close to the actual solution, in contrast to the points on the decoupled branch which could be much further away.
Thanks to the excellent convergence of the coupled re-iterations, however, the computational effort per solution point is not much larger than in the case of the coupled continuation method.
As the coupled re-iterations can be run completely in parallel, the wall clock time is reduced from two weeks to only two days, while the total number of core hours is slightly larger (\tref{computationalcosts}).
In fact, the wall clock time of the coupled re-iteration is in the order of magnitude of the (generally less accurate) state-of-the-art methods, and the total number of core hours is smaller than in the case of the AIC matrix method (\tref{computationalcosts}).
\fig{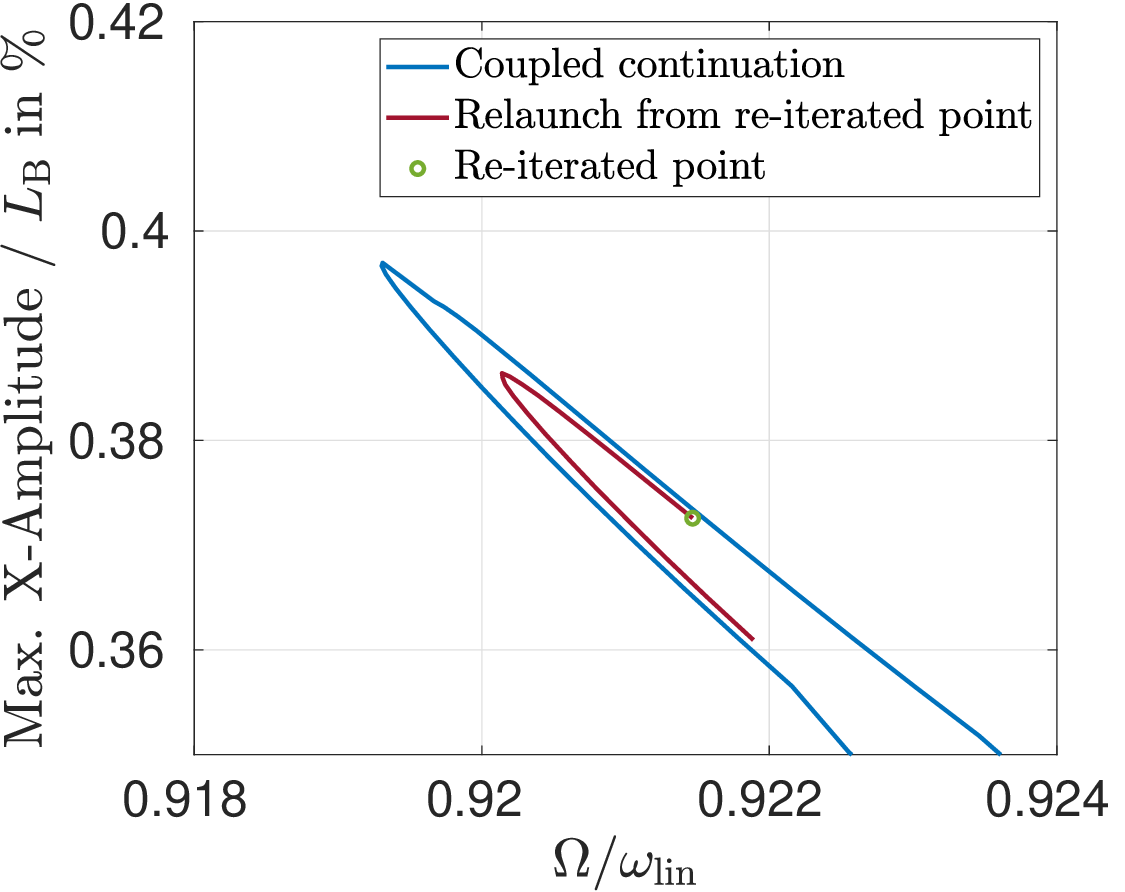}{
Amplitude-frequency curve near resonance peak: coupled continuation vs. relaunch of (decoupled) continuation from a re-iterated point
}{0.65}
\\
% RELAUNCH
A weakness of the coupled re-iteration is that the obtained points are not necessarily distributed in a favorable way along the new solution branch.
One means to close the gaps between the, in some regions scarce, re-iterated points is to relaunch path continuation form these points.
This, again, could be done in parallel for all re-iterated points.
The continuation should stop when another re-iterated point is reached.
One possibility is to carry out a fully coupled continuation when closing those gaps.
In \fref{relaunch_from_point_154}, instead, results are shown for a decoupled continuation.
More specifically, the approximation $\hat{\mm f}^*$ according to \eref{flin} is updated at the re-iterated point, and this approximation is subsequently used.
The important advantage of this procedure is that no further CFD simulation is needed, so that the computational effort is negligible.
The downside is that the results are slightly less accurate.
Indeed, the error increases with the distance from the re-iterated point.
Still, the amplitude deviation of the resonance peak is less than $3\%$, which can be deemed acceptable.
\section*{Summary and Conclusions}
The developed fully-coupled Harmonic Balance method is able to efficiently predict the forced response of turbomachinery blades accounting for both aerodynamic and structural mechanic (including contact) nonlinearities.
The coupling loop is implemented as a simple fixed-point algorithm and showed excellent numerical performance.
To improve convergence, the aerodynamic force within the structural solver is approximated as a consistent linear function, which exploits the fact that the vibration-induced force is in good approximation amplitude-linear in a sufficiently small neighborhood.
To robustly analyze the near-resonant response and to deal with turning points, the coupled solver was embedded in a numerical path continuation framework.
Two variants were developed, the coupled continuation of the solution branch and the coupled re-iteration of selected solution points.
In the former variant, the idea is to activate the coupling loop only when needed, and to use the aforementioned local approximation otherwise.
The important advantage of the coupled re-iteration is that it can be carried out completely in parallel, reducing the wall clock time to the order of magnitude of the current state-of-the-art methods which are based on aerodynamic influence coefficients.
The latter are generally less accurate, as they neglect the amplitude dependence of the aerodynamic forces and assume the linear superposition of wake- and vibration-induced flow, which was shown to have substantial effects on the predicted resonance amplitude in the numerical example.
To close the gap between the re-iterated points, it is proposed to relaunch the continuation procedure, again, in parallel, so that a sufficiently finely spaced solution path is retrieved.
\\
In the future, more sophisticated indicators could be developed that activate the coupling loop only when a significant deviation from linearity is encountered.
Moreover, the experimental validation of the developed prediction methods is planned.
\bibliographystyle{ieeetr}
\bibliography{literature_krack}

\end{document}